\documentclass[usenatbib]{mnras}

\usepackage{graphicx}	
\usepackage[fleqn]{amsmath}	
\usepackage{amssymb}	
\usepackage{multicol}        
\usepackage{bm}		
\usepackage{pdflscape}	

\usepackage[T1]{fontenc}
\usepackage{ae,aecompl}

\usepackage{newtxtext,newtxmath}

\usepackage[frozencache=true]{minted}		
\usepackage{enumitem,xcolor,xspace}
\usepackage{tikz}
\usepackage{booktabs}  

\colorlet{success}{green!60!green}
\usetikzlibrary{arrows, calc, positioning, patterns, shapes}

\definecolor{linkcolor}{rgb}{0, 0, 1.}			
\definecolor{funcolor}{rgb}{0.65, 0.16, 0.16}	
\definecolor{parcolor}{HTML}{1E8449}        
\definecolor{bg_code}{rgb}{0.95,0.95,0.95}		

\newminted{cpp}{fontsize=\footnotesize, gobble=2}
\newminted{python}{
mathescape=true, 
fontsize=\footnotesize, 
escapeinside=||, 
xleftmargin=0pt,
baselinestretch=1.2} 
\newminted{bash}{fontsize=\footnotesize} 

\newcommand{\tbref}[1]{Table~\ref{#1}}
\newcommand{\figref}[1]{Figure~\ref{#1}}
\newcommand{\secref}[1]{Section~\ref{sec:#1}}

\newcommand{\galario}{GALARIO\xspace}
\newcommand{\montblanc}{\texttt{montblanc}\xspace}

\newcommand{\uv}{\textit{uv}\xspace}

\newcommand{\py}{\texttt{Python}\xspace}
\newcommand{\CUDA}{\texttt{CUDA}\xspace}
\newcommand{\CC}{\texttt{C}\xspace}
\newcommand{\CXX}{\texttt{C++}\xspace}
\newcommand{\CUDACC}{\texttt{CUDA} \CXX}
\newcommand{\cufft}{\texttt{cuFFT}\xspace}
\newcommand{\cublas}{\texttt{cuBLAS}\xspace}
\newcommand{\fftw}{\texttt{FFTW}\xspace}
\newcommand{\openmp}{OpenMP\xspace}

\newcommand{\SI}[2]{#1\,\textrm{#2}}	
\newcommand{\githubrepo}{\href{http://github.com/mtazzari/galario}{http://github.com/mtazzari/galario}\xspace}
\newcommand{\githubdocs}{\href{https://mtazzari.github.io/galario/}{https://mtazzari.github.io/galario/}\xspace}
\newcommand{\githubissues}{\href{https://github.com/mtazzari/galario/issues}{https://github.com/mtazzari/galario/issues}\xspace}

\renewcommand{\vec}[1]{\mathbf{#1}}
\newcommand{\de}{\mathrm{d}}
\renewcommand{\Re}{\operatorname{Re}}	

\hypersetup{
colorlinks=true,
linkcolor=linkcolor,
citecolor=linkcolor,
filecolor=linkcolor,
urlcolor=linkcolor}

\newcommand{\func}[1]{\textcolor{funcolor}{\texttt{#1}}}	
\newcommand{\comm}[1]{\texttt{#1}}							

\hypersetup{
pdfauthor={M. Tazzari},
pdftitle={GALARIO: a GPU Accelerated Library for Analysing Radio Interferometer Observations},
pdfkeywords={gpu, radio astronomy, interferometry},
bookmarksnumbered=true,
draft=false}

\graphicspath{{./figures/}}

\title[GALARIO: a GPU Library for Interferometric Observations]{GALARIO: a GPU Accelerated Library for Analysing Radio Interferometer Observations}

\author[M. Tazzari, F. Beaujean and L. Testi]{Marco Tazzari$^{1}$\thanks{Contact e-mail: \href{mailto:mtazzari@ast.cam.ac.uk}{mtazzari@ast.cam.ac.uk}},
Frederik Beaujean$^{2}$ and
Leonardo Testi$^{2, 3}$
\\
$^{1}$Institute of Astronomy, University of Cambridge, Madingley Road, CB3 0HA,  Cambridge, UK\\
$^{2}$C2PAP, Excellence Cluster Universe, Ludwig-Maximilians-Universit\"at M\"unchen, Boltzmannstr. 2, D-85748 Garching, Germany\\
$^{3}$European Southern Observatory, Karl-Schwarzschild-Str. 2, D-85748 Garching, Germany}

\date{\today}

\pubyear{2018}

\begin{document}
\label{firstpage}
\pagerange{\pageref{firstpage}--\pageref{lastpage}}
\maketitle

\begin{abstract}
We present \galario, a computational library that exploits the power of modern graphical processing units (GPUs) to accelerate the analysis of observations from radio interferometers like ALMA or {the VLA}. \galario speeds up the computation of synthetic visibilities from a generic 2D model image or a radial brightness profile (for axisymmetric sources). On a GPU, \galario is 150 faster than standard Python and 10 times faster than serial \CXX code on a CPU.
Highly modular, easy to use and to adopt in existing code, \galario  comes as two compiled libraries, one for Nvidia GPUs and one for multicore CPUs, where both have the same functions with identical interfaces. \galario comes with Python bindings but can also be directly used in C or C++.
The versatility and the speed of \galario open new analysis pathways that otherwise would be prohibitively time consuming, e.g. fitting high resolution observations of large number of objects, or entire spectral cubes of molecular gas emission. It is a general tool that can be applied to any field that uses radio interferometer observations.
The source code is available online at \githubrepo under the open source GNU Lesser General Public License v3.
\end{abstract}




\section{Introduction}
In the quest for high angular resolution and high sensitivity, radio astronomy has been developing the use of interferometry since the late 1940s. 
Unlike single dishes, which directly measure the sky brightness and produce an image of it, radio interferometers measure \textit{visibilities}, the complex valued samples of the Fourier transform of the sky brightness \citep{Thompson:1999yu}. The locations in the Fourier plane where these samples are taken is {given} by the spatial distribution of the antennas on the ground {and the direction of the source being observed}.
Modern interferometers like Atacama large millimeter and sub-millimeter array (ALMA) and the {Karl G. Jansky Very Large Array (VLA)} have developed advanced pipelines that not only calibrate the observed visibilities, but also produce for the end users spectrally resolved images of the sky brightness distribution. 

Comparing a model prediction to an interferometric data set is typically done in one of the following two ways: either in the image plane by comparing a model image to the image of the sky reconstructed from the visibilities, or in the Fourier plane by directly comparing the observed visibilities to synthetic ones computed from the model image. The first approach is more intuitive but it is intrinsically limited: it relies on estimating the true sky brightness distribution from the observed visibilities. Unfortunately, the observations can only provide a finite number of samples of the visibilities, implying that a unique reconstruction of the sky brightness is not possible. In addition, to remove the effects of discrete sampling, non-linear deconvolution algorithms (e.g., the traditional CLEAN by \citealt{Hogbom:1974aa,Clark:1980lr} or MEM by \citealt{Cornwell:1985aa}) are applied to perform image reconstruction, which may introduce a variety of artefacts. Moreover, while each of the observed visibility point has associated a well behaved Gaussian noise (with equal variance in the real and imaginary part), the pixels in the reconstructed image have correlated noise whose properties are poorly constrained (due to the non-linear reconstruction algorithms). Ultimately, model comparison to the reconstructed images is thus affected in the image plane by the sampling of the sky visibility, the non-linear algorithms applied, and the correlated noise on the images, which reflects in the difficulty to correctly estimate the observational uncertainty \citep{Cornwell:1999aa}. The second approach -- comparing observed to model visibilities --
is much more straightforward as it operates in the domain where the observations were made and the uncertainties are better understood \citep{Pearson:1999aa}.

Comparing a model image computed on a regular grid to observed visibilities that are scattered across the Fourier plane involves a series of 1D and 2D array operations  such as Fourier transforms, transpositions, and interpolations \citep{Briggs:1999kq}. The size of  the arrays used to properly model the visibilities are set by the properties of the interferometer at observation time: spatial and spectral resolution, sensitivity, number and distribution of antennas. ALMA and the VLA have delivered tremendous improvements in terms of longer baselines, higher sensitivity, and more uniform Fourier plane coverage. This implies that the spectral and spatial sampling of the visibilities has increased enormously.

Inferring a model from the observations -- either using a Bayesian Markov chain sampler or a classical $\chi^2$ optimizer -- requires an adequate exploration of the parameter space. Performing the inference in the Fourier plane requires the computation of synthetic visibilities from the model image in each likelihood evaluation.
The enhancements in the quality of the visibilities delivered by modern interferometers has increased the computational effort required to model them accordingly, at a point where modeling medium-resolution observations can take one or more days of multi-core computation\footnote{A representative fit of a single wavelength continuum map at 0.1\arcsec resolution, assuming 4096x4096 matrix size, $10^6$ visibilities and $0.5$s to compute them with a standard \py code, $10^6$ likelihood evaluations to achieve convergence, running on 32 CPU cores, needs 49 wall-clock hours (excluding the model computation).}.

{The large software packages designed for the calibration and the management of interferometric datasets -- e.g., CASA\footnote{\href{https://casa.nrao.edu}{https://casa.nrao.edu}}, AIPS\footnote{\href{http://www.aips.nrao.edu}{http://www.aips.nrao.edu}}, MIRIAD \citep{Sault:1995aa}, GILDAS\footnote{\href{http://www.iram.fr/IRAMFR/GILDAS}{http://www.iram.fr/IRAMFR/GILDAS}} -- usually offer dedicated tasks for modelling the visibilities. However, although handy for a first characterisation of the observed sources, these tasks are often very limited in terms of flexibility: they typically require the user to choose among a very restricted set of simplified models for the source brightness, do not allow the user to specify what statistics should be used for the exploration of the parameter space, and cannot be easily incorporated into external modelling codes without a heavy performance penalty. In this context, the CASA-based UVMULTIFIT library \citep{Marti-Vidal:2014rf} constitutes a more flexible solution as it allows the user to model the visibilities with an indefinite number of parametric source components that can be personalised. We note that all the codes named so far are purely designed for CPUs, and only a few of them (e.g., CASA and UVMULTIFIT) can benefit from multi-core operations.}

A breakthrough in the computing capabilities is needed in order to fully and timely exploit the wealth of information that the new interferometers make available. In this paper we present \galario, a computational library that provides the necessary speed-up. Unlike the central processing units (CPUs) that are composed of at most a few tens of cores, the graphical processing units (GPUs) have thousands of cores that, although less powerful than CPU cores, effectively outperform CPUs in embarrassingly parallel tasks \citep{Nickolls:2008:SPP:1365490.1365500}, of which the operations needed to compute the synthetic visibilities  are eminent examples.

\galario is a library that uses GPUs or alternatively multiple CPU cores to speed up the computation of synthetic visibilities from a model image, and has been designed to achieve the best performance and still to be easy to use and to adopt in existing code. In the context of a fit, \galario can be easily adopted as a drop-in replacement to accelerate the computation of the $\chi^2$ between the model predictions and the observed visibilities. Moreover, thanks to its modular structure, \galario can be included in any likelihood computation, leaving to the user the choice of the statistical tool used for the parameter space exploration. The GPU version of \galario is about 150 times faster than standard \py implementations that rely on the widely used \comm{scipy} and \comm{numpy} packages, and ten times faster than serial C code.
From the user perspective, \galario can be called directly in \CC or \CXX and easily imported in \py code as a normal package.

To our knowledge, there is only another code, \montblanc \citep{Perkins:2015aa} that exploits the power of GPUs to compare models directly to observed visibilities. However, \galario differs from \montblanc in many aspects. First, \montblanc models the source brightness only through parametrised models (e.g., a point source, or a Gaussian ellipse) and does not support, as yet, unparametrised radio sources. Instead, \galario allows the user to compute synthetic visibilities from a generic 2D image of the sky brightness that can be the result, e.g., of a complex radiative transfer computation as well as of a simple parametric profile. Second, while \montblanc is dedicated to GPUs, all the functions of \galario are implemented both for GPU and CPU, on which the acceleration is achieved with OpenMP. Moreover, since the GPU and CPU functions in \galario have the same interfaces, it is easy to write reusable code that can be executed on the GPU or on the CPU just by changing which library is linked in (\CC) or imported (\py). 

The contexts in which \galario can be used are manifold. Originally developed in the field of protoplanetary discs, \galario implements a general computation of the synthetic visibilities {that makes} it suitable for application in any field dealing with observations from radio interferometers{for a wide range of wavelengths and angular resolutions}. 

{\galario has already been used in a few studies} to fit {moderate- and high-resolution} observations of protoplanetary discs. In \citet{Testi:2016aa} and \citet{Tazzari:2017aa} \galario was used to fit the visibilities of the disc continuum emission with a physical model to characterize the disc structure.  \citet{Tazzari:2016qy} used \galario to study the properties of the dust grains through the simultaneous fit of visibilities at multiple sub-mm, mm and cm wavelengths . In the domain of extreme high resolution observations \galario has been used to characterise the shape of the multiple rings appearing in the continuum emission of the AS~209 protoplanetary disk seen by ALMA \citep{Fedele:2017aa}.
It is worth noting that the speed-up that \galario delivers {naturally translates into the capability to extend the visibility analysis to} many objects on much reduced time scales, thus making it ideal to fit surveys of many sources \citep[e.g., {as has been done in}][]{Tazzari:2017aa}. Furthermore, a new pathway opened by the acceleration of \galario is the possibility to fit simultaneously entire spectral cubes of molecular-gas emission, allowing the kinematics of the object -- a protoplanetary disc or a galaxy -- to be characterised consistently.

The paper is organized as follows. Section~\ref{sec:overview} provides an overview of the code illustrating key functionalities and relevant use cases. Section~\ref{sec:theory} introduces the theoretical definitions and equations of Synthesis Imaging {and discusses the limitations of the current release of} \galario. Section~\ref{sec:implementation} describes the CPU and the GPU implementation of \galario and Section~\ref{sec:accuracy} presents the results of accuracy checks. In Section~\ref{sec:performance} we analyse the performance of \galario and in Section~\ref{sec:conclusions} we draw our conclusions. Appendix~\ref{app:install} summarizes the steps needed to obtain and install \galario. Appendix~\ref{app:performance} reports additional performance tests analogous to those discussed in Section~\ref{sec:performance}. {Appendix~\ref{app:accuracy} shows the results of additional accuracy checks carried out against the CASA package. Appendix~\ref{app:python.functions} presents the Python implementation of some reference functions.}

\section{Code overview}
\label{sec:overview}
In this Section we aim to give a quick overview of \galario for typical use cases that the reader might find immediately useful, deferring to Sect.~\ref{sec:theory} the definition of the quantities and equations involved.
\galario has been designed to accelerate the fundamental task of comparing a model prediction {(fundamentally, a brightness image)} with an interferometric observational data set, which typically consists of a collection of complex visibilities {$V_k$ ($k=1..M$) defined as the samples of the source visibility $V$ in discrete locations $(u_k,v_k)$}. The key functionality of \galario is the computation of synthetic visibilities given (i) a model image (or a radial brightness profile) and (ii) a collection of \uv-points {$(u_k,v_k)$}  representing the interferometric baselines sampled by the observations.

The core of \galario is written in \CXX (for the CPU version) and \CUDACC (for the GPU version). This allows \galario to achieve the best performances and to offer the same core functionalities in both versions.
The Python wrappers written in \comm{cython} are available for the main functions to facilitate the adoption of \galario in existing code.

On machines where no GPU is available, \galario can still provide a speed-up through \openmp on multiple CPU cores. If compiled and executed on machines with a CUDA enabled
GPU\footnote{The updated list of CUDA enabled GPUs is available at \href{https://developer.nvidia.com/cuda-gpus}{https://developer.nvidia.com/cuda-gpus}}, \galario delivers a dramatic speed-up with respect to normal CPU code, up to 150 times faster than a standard Python implementation that uses the \comm{numpy} and \comm{scipy} packages (more details in Sect.~\ref{sec:performance}).

\subsection{Selection of the version}

Both the CPU and GPU versions of \galario are compiled in single and double
precision. After installation, the CPU and GPU versions can be imported in
Python with
\begin{pythoncode}
from galario import double       # CPU
from galario import double_cuda  # GPU
\end{pythoncode}
The single- and double-precision libraries in both the CPU and GPU versions offer
the same functions with identical interfaces, thus making it easy to write
reusable code. Our recommended default is double precision. To use the
single-precision versions, replace \texttt{double} $\to$ \texttt{single} in the
above commands. The functions described below can be imported from any of these
four libraries.

\subsection{Basic usage}
The computation of the synthetic visibilities $V_\mathrm{mod}$ of a model image, {sampled} at some \uv-points {$(u_k,v_k)$} can be done with \func{sampleImage}:
\begin{pythoncode}
from double_cuda import |\func{sampleImage}|
Vmod = |\func{sampleImage}|(image, dxy, u, v)
\end{pythoncode}
where the \comm{image} is {a 2d array in} Jy/pixel units and its coordinate system is the same as that of the sky (East to the left, North to the top), \comm{dxy} is the size (in radians) of the image pixel {(assumed square)}, {\comm{u} and \comm{v} are linear arrays containing the coordinates $u_k$, $v_k$ (expressed in units of the observing wavelength $\lambda$),} and the returned array \comm{Vmod} is a complex array containing the synthetic visibilities (in Jy).

\begin{figure*}
\resizebox{\hsize}{!}{\includegraphics{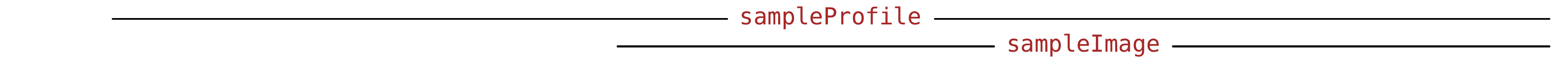}}
\resizebox{\hsize}{!}{\includegraphics{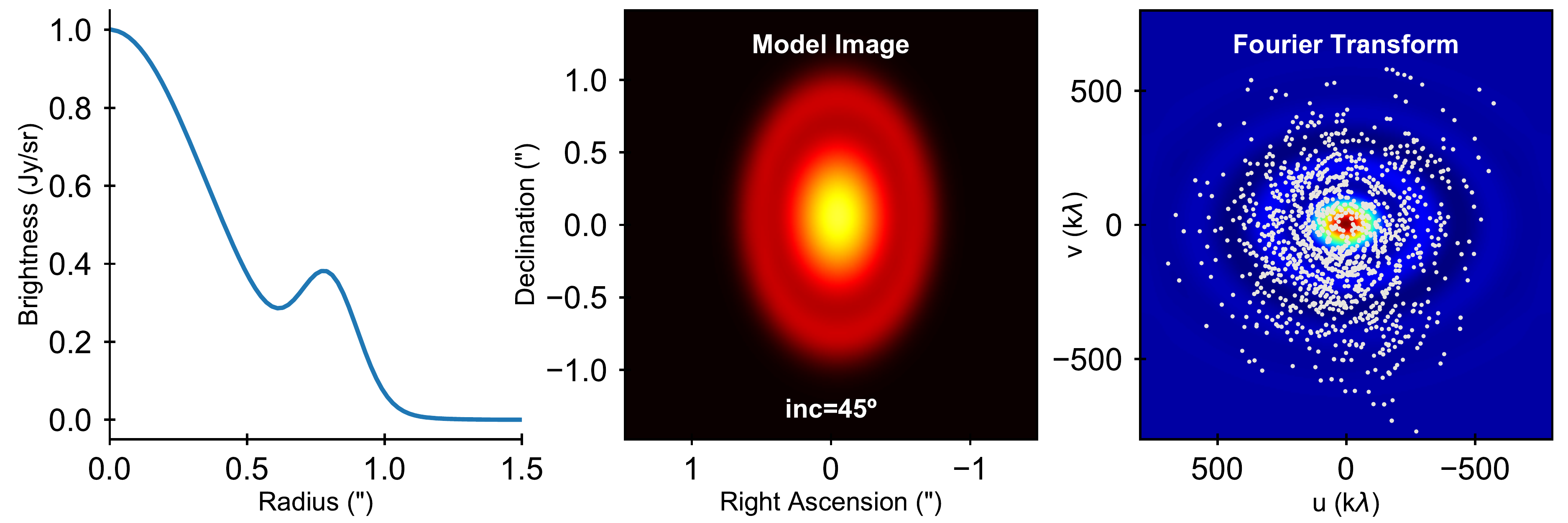}}
\caption{Workflow of the \func{sampleProfile} and \func{sampleImage} functions. \func{sampleImage} takes in input a 2D model image (central panel) and produces the synthetic visibilities by sampling its Fourier transform at the specified \uv-points locations (right panel). \func{sampleProfile} computes the synthetic visibilities in the same way as \func{sampleImage}, but takes in input a radial brightness profile (left panel) from which the model image is internally computed assuming axisymmetry and a line-of-sight inclination (45$^\circ$ in this example).}
\label{fig:code.overview}
\end{figure*}

\func{sampleImage} makes no assumptions on the symmetry of the 2D input \comm{image} and therefore can be used to compute the visibilities of any image.
However, in case the model image has an axisymmetric brightness distribution, \galario offers a faster version of \func{sampleImage} called \func{sampleProfile} that exploits the symmetry of the image and takes as input the brightness profile $I_\nu(r)$ defined on a radial grid and computes internally the 2D image by azimuthally sweeping the profile over $2\pi$:
\begin{pythoncode}
from double_cuda import |\func{sampleProfile}|
Vmod = |\func{sampleProfile}|(I, Rmin, dR, Nxy, dxy, u, v)
\end{pythoncode}
where {\comm{I} is a 1d array containing the radial brightness profile $I_\nu(R)$ (in Jy/sr),} \comm{Rmin} and \comm{dR} are the innermost radius and the cell size of the radial grid expressed in radians, and \comm{Nxy} is the number of pixels on each image axis. Figure~\ref{fig:code.overview} summarizes the workflow of \func{sampleProfile}: the radial brightness profile (left panel) is used to produce an axisymmetric 2D image (central panel) which is then Fourier transformed and sampled in the specified \uv-points (right panel).

The instruction above produces a face-on 2D image out of the profile $I_\nu(r)$. Producing an image with an inclination \comm{inc} (radians) along the line of sight can be done by specifying the optional parameter \comm{inc}:
\begin{pythoncode}
Vmod = |\func{sampleProfile}|(I, Rmin, dR, Nxy, dxy, u, v, inc=inc)
\end{pythoncode}
as shown in the example in \figref{fig:code.overview} for an inclination of $45^\circ$.

In the context of a fit, \galario provides handy functions to compute directly the likelihood of the model in terms of a $\chi^2$, both in the case the input is a model image or an axisymmetric brightness profile:
\begin{pythoncode}
chi2 = |\func{chi2Image}|(image, dxy, u, v, Re_Vobs, Im_Vobs, w)
chi2 = |\func{chi2Profile}|(I, Rmin, dR, Nxy, dxy, u, v,
                   Re_Vobs, Im_Vobs, w)
\end{pythoncode}
where \comm{Re\_Vobs}, \comm{Im\_Vobs} are the real and imaginary part of the observed visibilities and \comm{w} their {associated} weights.

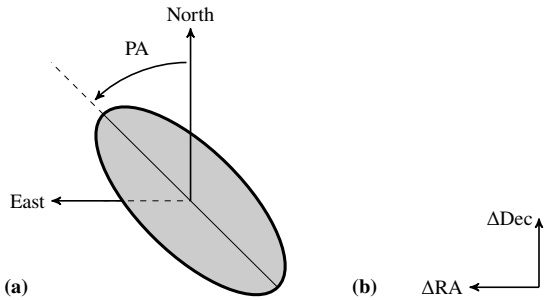
\begin{figure}
\begin{center}
\resizebox{0.85\columnwidth}{!}{\begin{tikzpicture}[
    scale=3,
    axis/.style={thick, ->, >=stealth'},
    arc/.style={thick, ->, >=stealth', shorten <=2pt, shorten >=2pt},
    dashed line/.style={dashed, thin},
    every node/.style={color=black}
    ]

\tikzset{
  font={\fontsize{10pt}{12}\selectfont}}
\draw[axis]  (1,0) -- (0.2,0) node[left] {East};
\draw[thin, dashed] (0.5, 0.5) -- (0.2, 0.8);

\draw[arc] (1,0.8) arc [start angle=90, end angle=135, radius=0.8] node[midway, above=0.2cm] {PA};

\filldraw[fill=gray!40!white, draw=black, line width=1.5]  (1,0) circle [x radius=0.707, y radius=0.3, rotate=-45];
\draw[axis] (1,0) -- (1,1) node[above] {North};
\draw[very thin] (1.5, -0.5) -- (0.5, 0.5);
\draw[very thin, dashed] (0.5, 0.) -- (1, 0.);

\draw[axis]  (3,-0.5) -- +(-.4,0) node[left] {$\Delta$RA};
\draw[axis]  (3,-0.5) -- +(0.,0.4) node[left] {$\Delta$Dec};

\draw (0., -0.5) node {\textbf{(a)}};
\draw (2., -0.5) node {\textbf{(b)}};
\end{tikzpicture}

}
\end{center}
\caption{Conventions used in \galario. (a) Definition of inclination and position angle (PA). A circular disc is inclined by $55^\circ$ and rotated by the position angle PA. The inclination is performed with a tilt along the North-South axis before rotating by PA (see, e.g., the central panel in \figref{fig:code.overview}). PA is the angle between the North-South axis and the line of nodes - the intersection of the plane of the object with the North-East plane - and is measured counter-clockwise (East of North). (b) Definition of the angular offsets. $\Delta$RA and $\Delta$Dec are positive for offsets towards East and North, respectively.}
\label{fig:def.PA.offsets}
\end{figure}

All the functions described so far support optional parameters useful to rotate and translate the model image given. It is possible to rotate the model image by a position angle PA, and to translate it by angular offsets in Right Ascension and Declination direction $(\Delta\mathrm{RA}, \Delta\mathrm{Dec})$ by specifying the optional parameters:
\begin{pythoncode}
Vmod = |\func{sampleImage}|(image, dxy, u, v,
                            PA=PA, dRA=|$\Delta$RA|, dDec=|$\Delta$Dec|)
\end{pythoncode}
where PA, $(\Delta\mathrm{RA}, \Delta\mathrm{Dec})$ are all expressed in radians and the offsets are defined in sky coordinates, i.e. positive $\Delta\mathrm{RA}$ and $\Delta\mathrm{Dec}$ translate the image towards East and North, respectively. Figure~\ref{fig:def.PA.offsets} illustrates these definitions.
The same optional parameters can be specified in \func{sampleProfile},  \func{chi2Image}, and \func{chi2Profile}. As described in Sect.~\ref{sec:theory.rotation.translation}, to achieve better performances the image rotation and translation are not applied to the model image but to the synthetic visibilities.

Hereafter we will refer to the \func{sampleProfile} and \func{sampleImage} functions by \func{sample*}, and analogously by \func{chi2*} for the \func{chi2Profile} and \func{chi2Image} functions. In a similar way, we will use \func{*Profile} and  \func{*Image} to indicate the related functions.

Details on how to install \galario are given in Appendix~\ref{app:install} and, more thoroughly, in the online documentation at \githubdocs, which also contains code examples showing how  \galario can be used in typical data analysis workflows.

\section{Visibility modelling}
The response of a synthesis array like ALMA and the VLA to the brightness distribution of a source in the sky is a collection of measurements called \textit{complex visibilities}. In this Section we introduce the basic equations needed to define the \textit{visibility} (a more thorough derivation can be found in \citealt{Wilson:2013aa}), {we illustrate how they can be implemented in a computer code, and we discuss the use cases and the limitations that follow from the adopted assumptions.}

\subsection{Basic equations of Synthesis Imaging}
\label{sec:theory}
To define the \textit{visibility} measurement, we first derive the response of a two-element interferometer, the fundamental receiving unit of the array. It consists of a correlator that combines, or multiplies and time averages, the signals received by the two antennas. A diagram of a two-element interferometer can be found in Figure~2-1 in \citet{Thompson:1999yu}. 

Let us introduce some definitions and a system of coordinates, following standard conventions as in \citet{Thompson:1999yu}.
Let $I_\nu(\vec s)$ be the source brightness in direction $\vec s$ at frequency $\nu$. $I_\nu(\vec s)$ is a {spectral brightness} and is measured in erg\,s$^{-1}$cm$^{-2}$Hz$^{-1}$sr$^{-1}$ or Jy\,sr$^{-1}$.
Let us assume the two antennas are identical, with response pattern $A(\bm \sigma)$ defined as the effective collecting area in direction $\vec s$. The radiation power collected from each of the antennas in direction $\vec s$ and received from the source element $\de\Omega$ in the frequency range $\Delta\nu$ is then $I_\nu(\vec s)A(\vec s) \Delta\nu \de\Omega$.
Let us call $\vec b$ the baseline vector connecting the two antennas on the ground and $\vec s$ the unit vector -- identical for both the antennas -- pointing towards the source. Under the simplifying assumption that the source brightness extends over a small region of the celestial sphere \citep{Clark:1999aa}, it is useful to rewrite $\vec s = \vec s_{0} + \bm \sigma$, where $\vec s_{0}$ is a unit vector representing the \textit{phase centre} of the synthesized field of view and $|\bm \sigma|\ll 1$. As a result, $\bm \sigma$, which is perpendicular to $\vec s_0$, lies in the plane tangent to the celestial sphere in $\vec s_0$.

Assuming that the source is in the far field of the interferometer {(the incoming wave fronts are plane parallel)} and its emission is incoherent {(different parts of the source emit uncorrelated radiation)},
it can be shown \citep{Clark:1999aa,Thompson:1999yu} that the response of a two-element interferometer to a source of brightness $I_\nu(\vec s)$ is
\begin{equation}
V(\vec b) = \int_{\Omega_{S}}
 \mathcal{A}(\bm \sigma) I_\nu (\bm \sigma)\, e^{{-2\pi i \nu\,{\vec b \cdot \bm \sigma}/{c}}}\mathrm d \Omega\,,
\label{eq:complex.visibility.obs}
\end{equation}
where $\Omega_{S}$ is the angular size of the source and $\mathcal{A}(\bm\sigma)=A(\bm \sigma)/A_0$ is the normalized antenna response pattern, with $A_0$ being the antenna response at the centre of the beam. {The central Gaussian-like feature of the antenna pattern is usually termed primary beam and is characterised by a full width at half maximum
\begin{equation}
\theta_{\mathrm{FWHM}} = K \frac{\lambda}{D}\,,
\label{eq:def.pb.hpbw}
\end{equation}}
where $D$ is the antenna diameter and $k$ is a numerical factor {close to unity. For reference, ALMA antennas have a measured $K=1.13$} \citep{alma-technical-handbook}. {The primary-beam full width at half maximum serves as the \textit{field of view} of single-pointing observations.}
{Eq.~\eqref{eq:complex.visibility.obs} --- derived assuming a bandwidth $\Delta \nu$ small enough so that $I_\nu$ and $\mathcal{A}$ can be considered effectively constant with $\nu$ ---} defines the \textit{complex visibility} of the source with respect to the chosen phase centre $\vec s_0$.

{In order to express Eq.~\eqref{eq:complex.visibility.obs} in a practical form}, it is useful to define a system of coordinates {such that the baseline vector $\vec b$ has coordinates $(u, v, w)$ where $u$ points towards the East, $v$ towards the North, and $w$ is parallel to the direction of interest (i.e., $\vec s_0$, the phase centre).} The coordinates $(u, v, w)$ are measured in units of the observing wavelength $\lambda=c/\nu_0$ {, with $\nu_0$ measured at the centre of the bandwidth.} {We also introduce a coordinate system on the sky $(l,m,n)$ with its origin in the phase centre and with $(l,m,n)$ being the direction cosines with respect to $u$ and $v$ such that}
\begin{equation}
\frac{\vec b\cdot \vec s}{\lambda} = ul+ vm+ wn\,.
\end{equation}
The $(l,m)$ plane is usually called \textit{image} plane because it is the plane on which the source brightness $I_\nu(l, m)$ is defined. {We note that inside the code the $(l, m)$ coordinates are termed $(x, y)$ to ease readability.} With these definitions we can rewrite the complex visibility \eqref{eq:complex.visibility.obs} as
\begin{equation}
V(u, v) =
\int\displaylimits_{-\infty}^{+\infty}
\int\displaylimits_{-\infty}^{+\infty}
\frac{\mathcal{A}(l, m) I_\nu(l, m)}{\sqrt{1-l^2-m^2}} e^{- 2\pi i (ul+vm+w\sqrt{1-l^2-m^2}-1)} \de l\, \de m\,.
\end{equation}
{Following \cite{Thompson:1999yu}, for small-field imaging, i.e. \linebreak $|(l^2+m^2)w|\ll 1$, the above expression simplifies to}
\begin{equation}
V(u, v) =
\int\displaylimits_{-\infty}^{+\infty}
\int\displaylimits_{-\infty}^{+\infty}
\mathcal{A}(l, m) I_\nu(l, m) e^{- 2\pi i (ul+vm)} \de l\, \de m\,,
\label{eq:complex.visibility.obs.xy}
\end{equation}
where the ranges of the integrals have been extended to infinity since the integrand $\mathcal{A} I_\nu$ is expected to be zero for $l^{2}+m^{2} > 1$. {Under the small-field imaging assumption,} Eq.~\eqref{eq:complex.visibility.obs.xy} shows that the  visibility $V$ of a source {of} brightness $I_\nu$ is the {two-dimensional} Fourier transform of its modified brightness distribution $\mathcal{A}I_\nu$. 

{For arrays with non-coplanar baselines ($w\neq 0$), the small-field imaging assumption introduces a phase error $\pi(l^2+m^2)w$ for radiation coming from the $(l,m)$ direction. \citet{Thompson:1999yu} and \citet{Cornwell:2008aa} show that this error is small in the region of the image plane centred in $(l,m)=(0,0)$ with angular diameter 
\begin{equation}
\theta_\mathrm{F} \lesssim \frac{\sqrt{\theta_{\mathrm{res}}}}{3}\,,
\label{eq:wprojection.limit}
\end{equation}
where $\theta_{\mathrm{res}}$ is the full width at half maximum of the synthesized beam (expressed in radians). For a reference observation at a resolution $\theta_{\mathrm{res}}= 0.1\arcsec $, this corresponds to a region $\theta_\mathrm{F} \lesssim 48\arcsec $ in the image plane. If the field of view of the observations (Eq.~\ref{eq:def.pb.hpbw}) is smaller than $\theta_\mathrm{F}$, then the small-field imaging assumption will be valid for single-pointing observations. 
}

The complex-valued visibility function $V_\mathrm{obs}(u, v)$ is {defined everywhere in the $(u,v)$ plane but it is only} measured at the discrete locations $(u_{k}, v_{k})$
{that correspond to the projected baselines at the moment of observation. These sampling locations are} usually {termed} \uv-points.
{In more general terms, the visibility  measurements made by the interferometer can be written as 
\begin{equation}
\label{eq:sampling.applied}
V_\mathrm{obs}(u_{k}, v_{k}) = S\, V_\mathrm{obs}(u, v)\,,
\end{equation}
where $S(u,v)$ is the visibility \textit{sampling} function} defined as
\begin{equation}
\label{eq:sampling.function}
S(u,v) = \sum_{k=1}^M \delta(u-u_k, v-v_k)\,,
\end{equation}
{where $\delta$ is the Dirac delta distribution.
}

In order to compare a model prediction to some observed visibilities {$V_\mathrm{obs}(u_{k}, v_{k})$}, we need to compute the synthetic visibilities of the model brightness $I_{\nu\,\mathrm{mod}}$ using Eq.~\eqref{eq:complex.visibility.obs.xy}:
\begin{equation}
V_\mathrm{mod}(u, v) =
\int\displaylimits_{-\infty}^{+\infty}
\int\displaylimits_{-\infty}^{+\infty}
\mathcal{A}(l, m) I_{\nu\,\mathrm{mod}}(l, m) e^{- 2\pi i (ul+vm)} \de x\, \de y\,,
\label{eq:complex.visibility.mod.xy}
\end{equation}
and then sample $V_\mathrm{mod}$ at the same \uv-points where the observations were taken.
{The model likelihood, i.e. the probability of obtaining the observed data assuming the model is correct,} can be estimated by {means of a Gaussian likelihood \citep{Pearson:1999aa} $\mathcal{L}\propto \exp(-\chi^2/2)$ where}:
\begin{align}
\chi^{2}
&= \sum_{k=1}^{M} \chi^2_k = \sum_{k=1}^{M}  \left| V_{\mathrm{obs}}(u_{k}, v_{k})-V_{\mathrm{mod}}(u_{k}, v_{k}) \right|^{2} w_{k} \,\,,
\label{chap6.eq:def.chi.square}
\end{align}
where $w_{k}$ is the weight associated to the $k-$th observed visibility. The weights are computed theoretically as described in \citet{Wrobel:1999gf} {and should reflect the standard deviation $\sigma_k$ of the measurements of $V(u_k,v_k)$ such that $w_{k}=1/\sigma_k^2$. 
}

{
\subsection{Summary of the assumptions in the first release}
\label{sec:theory.assumptions}
In this Section we discuss some relevant assumptions made in the first released version of the code:
\begin{enumerate}
\item 
small-field imaging: the first release of \galario uses Eq.~\eqref{eq:complex.visibility.obs.xy} to compute the visibilities, thus neglecting the non-coplanarity of the baselines. This restricts the usage of the code to the cases in which the the region modelled with \func{*Image} or \func{*Profile} lies within the region defined in Eq.~\eqref{eq:wprojection.limit}.
\item 
Primary-beam correction: the \func{*Image} functions take as input an image of the primary-beam corrected brightness $ \mathcal{A}I_\nu(l,m)$. In the cases in which the region of interest in the image plane is small compared to the primary beam and close to its centre, one can approximate $\mathcal{A}I_\nu\approx I_\nu$ and apply the \func{*Image} functions directly to the brightness without significant deviations. The choice whether to apply this approximation is left to the user. We note, however, that in the first released version of the code
the \func{*Profile} functions --- which take as input a profile $I_\nu(R)$ and internally compute $I_\nu(l,m)$ --- do not apply the primary beam correction.  
\item 
Frequency dependence of $\mathcal{A}$ and $I_\nu$: both the antenna pattern and the source brightness are frequency-dependent quantities. As stated in the previous Section, the definition in Eq.~\eqref{eq:complex.visibility.obs} holds for small bandwidths $\Delta \nu$ over which the integrand can be assumed constant. For this reason, in the first release of \galario, the visibilities are assumed all at the same average frequency $\nu_0$. This implies that, in order to compare synthetic visibilities to observed ones (e.g. through Eq.~\eqref{chap6.eq:def.chi.square} with the \func{chi2*} functions), the observed visibilities (typically consisting of multiple measurements over several hundreds of spectral channels) must be channel-averaged\footnote{{This can be achieved, e.g., with the \comm{split} command of the Common Astronomy Software Application (CASA) package.}} into a single channel at frequency $\nu_0$ and characterised by a small $\Delta \nu$. We note that the effect of channel averaging is to combine the brightness measurements over a region with angular extent $\frac{\Delta\nu}{\nu_0}\sqrt{l^2+m^2}$ along the radial direction. Often termed \textit{bandwidth smearing}, this effect is not negligible at the  distances $\sqrt{l^2+m^2}$ where its angular extent becomes comparable with the synthesized beam.  The user can choose $\Delta\nu$ in order to control the bandwidth smearing within the image plane region of interest.
\end{enumerate}

The computation of synthetic visibilities of a field of view with multiple sources can be done in basically two ways: either by applying \func{*Image} to an image of $\mathcal{A}I_\nu(l,m)$ containing all the sources, or by summing up the visibilities of each single source computed independently with either \func{*Image} or \func{*Profile}.  In the second approach, the displacement of each source in the field of view can be achieved (at a small computational cost) by applying a different complex phase to the individual visibilities as described in the next Section. While the first approach requires executing only one Fourier transform --- appearing theoretically more computationally convenient --- the second approach exploits the linearity of the Fourier transform and might yield results faster if there are many identical sources to be placed in different locations. 

It is worth highlighting that in all cases (single or multiple sources in the field of view), the limitations due to the assumptions (i) to (iii) apply: all the sources must be located in a region that is close to the phase centre and small compared to $\theta_{\mathrm{F}}$ and the synthetic visibilities are computed in a narrow band around the observing frequency $\nu_0$.
}

\subsection{Image translation and rotation}
\label{sec:theory.rotation.translation}
The \func{*Profile} and \func{*Image} functions enable the user to apply a translation and a rotation with respect to the phase centre to the model image by specifying the optional parameters \comm{dRA}, \comm{dDec} and \comm{PA}. This functionality can be useful, e.g., to fit the centre and the Position Angle of a model image to the observations. Instead of translating and rotating the model image before taking the Fourier transform, \galario exploits the symmetries of the Fourier transform under these geometric operations to achieve a better performance and accuracy \citep{Briggs:1999kq}.

To perform the rotation, we use the fact that the Fourier transform commutes with rotations. This implies that to compute the visibilities of a model $I_{\nu\,\mathrm{mod}}$ rotated by an angle $PA$, it is sufficient to rotate the coordinates of the \uv-points by $-PA$ with
\begin{align}
u_k' &= u_k\cos(PA) - v_k\sin(PA)\\
v_k' &= u_k\sin(PA) + v_k\cos(PA)
\end{align}
where $u_k'$ and $v_k'$ are the rotated coordinates of the $k$-th \uv-point.

The translation of the model image is obtained by multiplying the sampled visibilities $V_\mathrm{mod}(u_k, v_k)$ by a complex phase, rather than by interpolating the image on a shifted spatial grid. This is based on the behaviour of the Fourier transform with respect to translations, according to which
\begin{equation}
\mathfrak{F}\, g(x-\Delta x) = \mathfrak{F}\, g(x) \times e^{-2\pi i u \Delta x }\,,
\end{equation}
where $\mathfrak{F}$ denotes the Fourier transform operation. By multiplying the sampled visibilities $V_\mathrm{mod}(u_k, v_k)$ by a phase $\exp[-2\pi i (u\Delta\alpha + v \Delta\delta)]$ with $u,\ v$ measured in units of wavelength and $\Delta\alpha,\ \Delta\delta$ measured in radians, it is possible to apply the desired shift in the image plane.

\subsection{Requirements on the image}
\label{sec:theory.requirements}
To compute the Fourier transform in Eq.\eqref{eq:complex.visibility.obs.xy} \galario uses the  fast Fourier transform algorithm (FFT) \citep{Cooley:1965:AMC} that requires  a regularly spaced 2D image as input. In this Section we describe the constraints on the image size and the pixel size that should be fulfilled for a correct computation of the complex visibilities. {We note that such constraints are jointly determined by the distribution of \uv-points (which sets the resolution and the maximum recoverable scale of the observations), by the diameter of the antennas (which sets the primary beam), and by the size and the location of the sources (which set the portion of the image plane of interest). For the clarity of the exposition, the considerations that follow are derived assuming a single source in a single-pointing observation. The generalisation for multiple sources is at the end of this Section.}

Let us call $N_l$ and $N_m$ the number of pixels in the $l$ and $m$ direction, respectively, of the input matrix containing $\mathcal{A}I_\nu(l,m)$.
The origin $(l, m)=(0, 0)$ is located at the image centre. 
If $\Delta\theta_l$ and $\Delta\theta_m$ are the angular pixel sizes in each direction, the input matrix covers a rectangular region in the image plane defined by
\begin{equation}
|l|\leq\frac{N_l\Delta\theta_l}{2}\quad \mathrm{and}\quad |m|\leq\frac{N_m\Delta\theta_m}{2}\,.
\end{equation}
In an analogous way, we can introduce the pixel size in the \uv-plane $\Delta u$ and $\Delta v$, in the $u$ and $v$ direction, respectively. The region of the \uv-plane covered by the output matrix of the FFT algorithm is thus defined by
\begin{equation}
|u|\leq\frac{N_l\Delta u}{2}\quad \mathrm{and}\quad |v|\leq\frac{N_m\Delta v}{2}\,.
\end{equation}
There is a correspondence between the pixel size in the image plane and that in the \uv-plane, given by
\begin{equation}
\label{eq:theory.def.fov}
N_l \Delta\theta_l = \frac{1}{\Delta u}\quad \mathrm{and} \quad
N_m \Delta\theta_m = \frac{1}{\Delta v}\,.
\end{equation}
In the remainder of this discussion, let us assume square pixels both in the image plane and in the \uv-plane:
\begin{equation}
\Delta\theta_l=\Delta \theta_m\equiv\Delta\theta_{lm}\quad \mathrm{and} \quad
\Delta u=\Delta v \equiv\Delta uv\,.
\end{equation}
This is a choice that is usually made and it is also assumed inside \galario. For the present discussion let us also assume for simplicity that the input matrix is square; i.e.,
\begin{equation}
N_l = N_m \equiv N_{lm}\,.
\end{equation}

The distribution of \uv-points where the synthetic visibilities have to be computed imposes two fundamental constraints on the values of $N_{lm}$, $\Delta\theta_{lm}$, and $\Delta uv$:
\begin{description}
\item (i)\quad
the region of the \uv-plane that is modelled must encompass the region sampled by the \uv-points, exceeding the most extended baseline by at least a factor of two in order to fulfil Nyquist sampling, that is:
\begin{equation}
\frac{N_{lm} \Delta uv}{2} = \max_{k}\{({u^{2}_{k}+v^{2}_{k}})^{1/2}\}\cdot f_\mathrm{max} 
\quad \mathrm{with}\quad f_\mathrm{max} > 2\,,
\end{equation}
where the maximum is taken over all the baselines represented by the given \uv-points.

\item (iia)\quad the region of the image plane that is modelled must be \textit{at least} larger than the maximum recoverable scale $\theta_{\mathrm{MRS}}$, namely:
\begin{equation}
\label{eq:theory.def.nxy}
N_{lm}\Delta \theta_{lm} > \theta_{\mathrm{MRS}}\equiv\frac{\Gamma}{\min_{k}\{({u^{2}_{k}+v^{2}_{k}})^{1/2}\}}\,,
\end{equation}
{where $\Gamma\approx 0.5$ is a constant. For reference, ALMA has $\Gamma=0.6$ \citep[cf. Eq. 3.27 in][]{alma-technical-handbook}}. Using Eq.~\eqref{eq:theory.def.fov} we can rewrite Eq.~\eqref{eq:theory.def.nxy} as a constraint on the \uv cell size:
\begin{equation}
\Delta uv = \frac{1}{\Gamma\,f_\mathrm{min}} \cdot \min_{k}\{({u^{2}_{k}+v^{2}_{k}})^{1/2}\}
\quad \mathrm{with}\quad 
f_\mathrm{min} > 1\,,
\end{equation}
{and thus compute the image size}:
\begin{equation}
\label{eq:theory.def.duv}
N_{lm}= 2 \Gamma f_\mathrm{min} \frac{ \max_{k}\{({u^{2}_{k}+v^{2}_{k}})^{1/2}\}}{\min_{k}\{({u^{2}_{k}+v^{2}_{k}})^{1/2}\}}\,.
\end{equation}
A conservative choice for $f_\mathrm{min}$ would be $f_\mathrm{min}=5$ to ensure that the field of view of the input matrix encompasses at least by five times the scale of the largest sources that might be resolved in the data.
\end{description}
{The most conservative criterion for the choice of $\Delta uv$ consists of imaging the whole field of view covered by the observations:} 

\noindent\ \ (iib)\quad the region of the image plane that is modelled must be as large as the primary beam, namely:
\begin{equation}
N_{lm}\Delta \theta_{lm} = \theta_{\mathrm{FWHM}}\,.
\end{equation}
{In this case, the image size is given by}
\begin{equation}
\label{eq:theory.def.duv.pb}
N_{lm}= 2 \frac{D}{K\lambda}\,\max_{k}\{({u^{2}_{k}+v^{2}_{k}})^{1/2}\}\,,
\end{equation}
{which typically yields much larger $N_{lm}$ than Eq.~\eqref{eq:theory.def.duv}.}

Given a distribution of \uv-points $(u_k, v_k)$, these criteria allow one to compute $N_{lm}$ and $\Delta\theta_{lm}$ {that should be used for} the input image. These criteria are implemented in \func{get\_image\_size}, which can be used as
\begin{pythoncode}
|$N_{lm}$, $d_{lm}$| = |\func{get\_image\_size}|(u, v, PB=|$\theta_\mathrm{FWHM}$|)
\end{pythoncode}
By default \func{get\_image\_size} uses criterion (iia) and Eq.~\eqref{eq:theory.def.duv}. If  the primary beam FWHM is specified as the optional parameter \comm{PB}, then \func{get\_image\_size} uses Eq.~\eqref{eq:theory.def.duv.pb} {in criterion (iib)}.

{In case the field of view contains multiple sources, criterion (iib) (instead of iia) should be used to ensure that the sources are correctly represented in the image plane. In any case, $N_{lm}$ should always be large enough so that the sources are far from the edges of the image. Finally, we note that  $N_{lm}$ is ultimately limited by the assumptions discussed in Section~\ref{sec:theory.assumptions}.}

Table~\ref{tb:array.config.matrix.properties} shows a compilation of matrix properties derived for realistic ALMA and {VLA} array configurations.
For each configuration we report the nominal minimum and maximum baseline, $N_{lm}$ computed using both criteria (iia) and (iib), $\Delta\theta_{lm}$ and the resolution $\theta_\mathrm{res}$. $\Delta\theta_{lm}$ and $\theta_\mathrm{res}$ depend on the observing wavelength, for which we assumed representative values of $\lambda=1.3$\,mm for ALMA and $\lambda=7.0$\,mm for {the VLA}.
In creating the Table, we computed $\theta_{\mathrm{res}}=(\mathrm{Max\ Baseline}\,\lambda^{-1}\Delta uv)^{-1}$, which is an ideal estimate that assumes a natural weighting scheme and neglects inhomogeneities in the baseline distribution; {real} values {depend on the actual distribution of the antennas and }differ at most by 15\% (cf. {\citealt{alma-technical-handbook}}).
We notice that the typical matrix sizes requested to cover the MRS by at least a factor of five ($f_\mathrm{min}=5$) range between 256$^2$ and 4096$^2$ for ALMA and between 512$^2$ and 2048$^2$ for {the VLA}; much larger matrix sizes (up to 16384$^2$) are needed to cover the full primary beam (we caveat that the image sizes $N_{lm}$ (iib) reported for the VLA A and B configurations exceed the small field imaging assumption). In all cases the values of $\Delta\theta_{lm}$ are comfortably smaller then the synthesized beam $\theta_\mathrm{res}$ by 5-10 times.
\begin{table}
\caption{Matrix and pixel sizes for ALMA and VLA configurations}
\begin{tabular}{lrrrrrr}
\toprule
	&	\multicolumn{2}{c}{Baselines}	&	\multicolumn{3}{c}{Matrix properties}		&	\\\cmidrule(lr){2-3}\cmidrule(lr){4-6}
Array	&	Min	&	Max	&	$N_{lm}$ (iia)	&	$N_{lm}$ (iib)	&	$\Delta\theta_{lm}$	&	$\theta_{\mathrm{res}}$\\
Config.		&	(m)	&	(m)	&	(px)	&	(px)	&	(\arcsec)	& (\arcsec) \\
\midrule
\multicolumn{7}{c}{ALMA} \\
C43-1	&	 14.6	&	160.7	&	256	&	256	&	0.215	&	1.669	\\
C43-2	&	 14.6	&	313.7	&	512	&	512	&	0.108	&	0.855	\\
C43-3	&	 14.6	&	500.2	&	1024	&	1024	&	0.054	&	0.536	\\
C43-4	&	 14.6	&	783.5	&	1024	&	1024	&	0.054	&	0.342	\\
C43-5	&	 14.6	&	1397.9	&	2048	&	2048	&	0.027	&	0.192	\\
C43-6	&	 14.6	&	2516.9	&	4096	&	4096	&	0.013	&	0.107	\\
C43-7	&	 64.0	&	3637.8	&	1024	&	4096	&	0.012	&	0.074	\\
C43-8	&	110.4	&	8547.7	&	2048	&	8192	&	0.004	&	0.031	\\
C43-9	&	367.6	&	13894.2	&	1024	&	16384	&	0.002	&	0.019	\\
C43-10	&	244.0	&	16194.0	&	1024	&	16384	&	0.003	&	0.017	\\[10pt]
\multicolumn{7}{c}{VLA} \\
A	&	680.0	&	36400.0	&	1024	&	16384	&	0.006	&	0.040	\\
B	&	210.0	&	11100.0	&	1024	&	4096	&	0.020	&	0.130	\\
C	&	 35.0	&	3400.0	&	2048	&	2048	&	0.060	&	0.425	\\
D	&	 35.0	&	1030.0	&	512	&	512	&	0.242	&	1.402	\\
\bottomrule
\end{tabular}
\begin{flushleft}{Notes}\quad The baselines are taken from the ALMA Cycle 5 Technical Handbook and the VLA 2018A Call for proposal. $N_{lm}$ (iia) and $N_{lm}$ (iib) have been computed using criteria (ii) and (iii) in Eq.~\eqref{eq:theory.def.duv} and Eq.~\eqref{eq:theory.def.duv.pb}, respectively. We used $f_{\mathrm{max}}=2.5$ and  $f_{\mathrm{min}}=5$. $\Delta\theta_{lm}$ and $\theta_{\mathrm{res}}$ have been computed assuming representative values $\lambda=1.3$\,mm for ALMA and $\lambda=7.0$\,mm for {the VLA}. We caveat that the image sizes $N_{lm}$ (iib) for the VLA A and B configurations exceed the maximum allowed by the small-field assumption.
\end{flushleft}
\label{tb:array.config.matrix.properties}
\end{table}%

For best performances in the FFT computation, it is advisable to use matrices with $N_{lm}$ that is a power of two.

The last step in the visibilities computation requires sampling the matrix containing $V(u,v)$  at the discrete locations $(u_k, v_k)$, {as described by Eq.~\eqref{eq:sampling.applied}}.
This operation can be done either by convolving $V(u,v)$ with a carefully chosen {kernel and then by sampling the result at the centre of each grid cell} \citep{Schwab:1984aa,Briggs:1999kq}, or by means of interpolation.
\galario performs the sampling using a bilinear interpolation algorithm \citep{Press:2007nr}; i.e., by inferring the value $V_\mathrm{mod}(u_k, v_k)$ from the value of $V(u,v)$ in the four closest grid points, assuming linear increments in both directions.

\section{Implementation}
\label{sec:implementation}

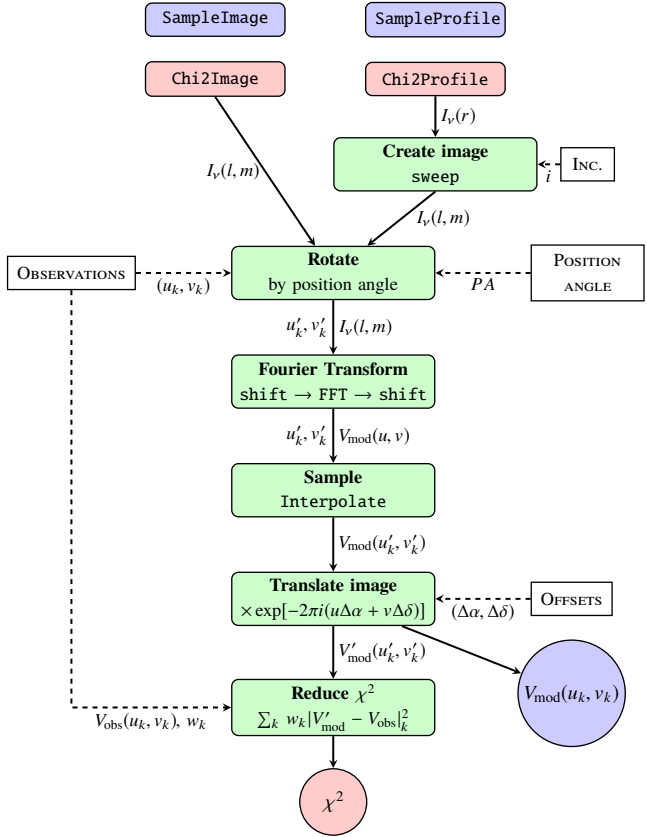
\begin{figure}
\centering
\resizebox{\columnwidth}{!}{\begin{tikzpicture}[node distance=2.3cm, >=stealth]
\tikzset{
  font={\fontsize{11pt}{12}\selectfont}}
  \tikzstyle{mst}=[thick, rectangle, draw=black, inner sep=0.2cm, 
  text width=7em, text centered]  
  \tikzstyle{mst-optional}=[dotted, thick, rectangle, draw=black, inner sep=0.2cm] 


    \tikzstyle{inputs} = [rectangle, draw, thick, text centered,
    inner sep=0.2cm, minimum width=4em, minimum height=2.5em]
  \tikzstyle{block} = [rectangle, draw, fill=success!20, thick,
  text width=14.5em, text centered, rounded corners, minimum height=4em]
  \tikzstyle{small-block} = [rectangle, draw, fill=success!20, thick,
  text width=6em, text centered, rounded corners, minimum height=2em]
  \tikzstyle{line} = [draw, -latex']
  \tikzstyle{cloud} = [draw, circle, thick, node distance=2cm,
  minimum height=2.5em, minimum width=2.5em]
  \tikzstyle{funcSample}=[thick, rectangle, draw, fill=blue!20, inner sep=0.2cm, rounded corners, text width=9em, text centered, minimum height=2.5em] 
    \tikzstyle{funcchi2}=[thick, rectangle, draw, fill=red!20, inner sep=0.2cm, rounded corners, text width=9em, text centered, minimum height=2.5em] 
  \tikzstyle{resultsample} = [draw, circle, thick, node distance=2cm, fill=blue!20,
  minimum height=5em, minimum width=5em]
    \tikzstyle{resultchi2} = [draw, circle, thick, node distance=2cm, fill=red!20,
  minimum height=5em, minimum width=5em]
  \matrix[column sep=1cm, row sep=0.5cm] (m) {
   \node[funcSample] (sampleimage) {\texttt{SampleImage}}; & \node[funcSample] (sampleprofile) {\texttt{SampleProfile}};\\
    \node[funcchi2] (chi2image) {\texttt{Chi2Image}}; & \node[funcchi2] (chi2profile) {\texttt{Chi2Profile}};\\[10pt]
    & \node[block] (sweep) {\textbf{Create image}\\[3pt]\texttt{sweep}};\\
  };

  \node[block, below of=sweep, anchor=east] (rot) {\textbf{Rotate}\\[3pt]by position angle};
  \node[block, below of=rot ] (ft) {\textbf{Fourier Transform}\\[3pt]\texttt{shift} $\to$ \texttt{FFT} $\to$ \texttt{shift}};
  \node[block, below of= ft] (sampl) {\textbf{Sample}\\[3pt]\texttt{Interpolate}};
  \node[block, below of=sampl] (transl) {\textbf{Translate image}\\[3pt]$\times\exp[-2\pi i (u\Delta\alpha + v \Delta\delta)]$};
  \node[block, below of=transl] (compar) {\textbf{Reduce $\chi^2$}\\[3pt]
    {$\sum_{k}\,w_k |V_{\mathrm{mod}}'-V_{\mathrm{obs}}|_k^2$}};
  \node[resultchi2, below of=compar] (chi) {\Large$\chi^2$};
  \node[inputs, right=2cm of transl] (offsets) {\textsc{Offsets}}; 
\node[resultsample, below of=offsets] (vmod) {\Large $V_{\mathrm{mod}}(u_k, v_k)$};  

  \draw[->,  very thick, dashed] (offsets) -- node[below] {$(\Delta \alpha, \Delta \delta)$} (transl);
  \node[inputs, left=2cm of rot] (ob) {\textsc{Observations}};
  \draw[->, very thick, dashed] (ob) |- node[below, near end] {$(u_k,v_k)$} (rot.west);
  \draw[->, very thick, dashed] (ob.south) |- node[below, near end] {$V_{\mathrm{obs}}(u_k,v_k),\, w_k$} (compar.west);
  \node[mst, right=2cm of rot] (pa) {\textsc{Position}\\[3pt]\textsc{angle}};
  \node[inputs, above of=pa] (inc) {\textsc{Inc.}};
    \draw[->, very thick, dashed] (inc) -- node[below] {$i$} (sweep);

  \draw[->, very thick, dashed] (pa) -- node[below] {$PA$} (rot);

  \draw[->, very thick, color=black] (chi2image.south) -- node[left] {$I_\nu(l,m)$} (rot);
  \draw[->, very thick, color=black] (chi2profile) -- node[right] {$I_\nu(r)$} (sweep);
  \draw[->, very thick, color=black] (sweep.south) -- node[right=2mm] {$I_\nu(l,m)$} (rot);
  \draw[->, very thick, color=black] (ft) -- node[left] {$ u'_k, v'_k$} node[right] {$ V_{\mathrm{mod}}(u, v)$} (sampl);
  \draw[->, very thick, color=black] (rot) -- node[left] {$ u'_k, v'_k$} node [right] {$I_\nu(l,m)$}(ft);
  \draw[->, very thick, color=black] (sampl) -- node[right] (){$ V_{\mathrm{mod}}(u'_k, v'_k)$} (transl);
  \draw[->, very thick, color=black] (transl) -- node[right] () {$ V'_{\mathrm{mod}}(u'_k, v'_k)$} (compar);
  \draw[->, very thick, color=black] (compar) -- node[left] {} (chi);
  \draw[->, very thick, color=black] (transl) -- node[left] {} (vmod);

\end{tikzpicture}

}
\caption{Flow chart of the algorithm, proceeding from top to bottom. White boxes indicate inputs, green boxes represent operations involving one or more parallel regions that can run on the GPU or the CPU. Circles indicate outputs and are colour-coded as the functions from which they are produced: red for the \func{chi2*} functions and the purple for the \func{sample*} functions. Arrows indicate data flow between kernels.}
\label{fig:flow}
\end{figure}

The basic purpose of \galario is to compute synthetic visibilities at
a set of points in the $uv$-plane as illustrated in
\figref{fig:code.overview}. To achieve this, a number of operations
have to be carried out. In \figref{fig:flow} 
we show the relevant
operations that are common to CPU and GPU as a flow chart in order to
compute the visibilities ($V_\mathrm{mod}$) and the $\chi^2$. The functions \func{Chi2Image} and \func{Chi2Profile}
only differ in the first stage, where the input image is either taken
as is or created from a radial profile. The next steps before the
$\chi^2$ reduction create the visibilities. If the users wishes to use
these directly, perhaps in a more sophisticated analysis than a
$\chi^2$ fit, then \func{sampleImage} and \func{sampleProfile} would
return at that point.

All operations shown in \figref{fig:flow} have a multi-threaded CPU and GPU
implementation. We wrote the code in \CXX and parallelized custom kernels in
 \cite{CUDA_guide} on the GPU, and with the help of
\openmp\citep{Dagum:1998:OIA:615255.615542} on the CPU. For custom kernels, we
used common inline functions to inject the core operations into surrounding code
that differs on CPU and GPU because of memory handling or available libraries.
GPU kernels use grid-stride loops when applicable. Whenever possible, we prefer
optimized library functions instead of custom kernels. The FFT is performed by
\fftw~\citep{FFTW05} or \cufft~\citep{cufft}. We use \cublas \citep{cublas} for
the $\chi^2$ reduction on the GPU.

To simplify the flow chart, we omit the memory operations because they
are quite different on CPU and GPU. We assume that, prior to calling
\galario, observations and all other inputs are initially in the CPU
main memory. To use the GPU, these data have to be transferred and
this can take a significant fraction of the overall execution time
whereas the transfer is unnecessary when computing on the CPU; see
\secref{performance} for details. In the special case of an axisymmetric brightness profile, we can exploit the symmetry of the image to avoid unnecessary data transfer: we supply the \func{*Profile} functions that only copy a radial profile defined on sky coordinates
and create the image directly in the GPU memory through the \func{sweep}
function, which essentially rotates the profile to sweep
the 2D image over $2\pi$, performing bilinear
interpolation as in \cite{Press:2007nr}. The purpose of the \func{Chi2*}
function is to avoid transferring the sampled visibilities back from
the GPU.

In the typical use case, an input image is such that the origin of the
coordinate system is in the central pixel. 
But \comm{FFTW} and
\comm{cuFFT} expect the origin in the top-left pixel. So we copy or
create the input image in a buffer and perform the shift, the FFT, and
the inverse shift in place. The shift algorithm is similar to the one
by \cite{Abdellah:2014} but was independently devised. The input image
is real which saves a factor of two in both memory and computing
effort in the FFT compared to a complex-to-complex transform.

All operations of \figref{fig:flow} are accessible separately, which greatly help with unit testing. We build up an extensive suite of tests using
\comm{pytest} that verifies the individual operations and their
various combinations. To improve the speed of \galario, we used
graphical profilers such as Nvidia \comm{nvvp} and Intel
\comm{Amplifier} as well as custom timing methods to
continuously monitor the performance in a more automated fashion.

\section{Accuracy}
\label{sec:accuracy}
In this Section we report the results of the tests that we conducted to check the accuracy  of \galario against analytic results. {In Appendix~\ref{app:accuracy} we report additional accuracy checks performed against the NRAO CASA package for input images that do not necessarily have analytic visibility expressions.} 

To {check the accuracy of \galario against analytic results}, we use the fact that the synthetic visibilities of an axisymmetric brightness profile $I_\nu(r)$ centred at the origin of the image plane have an analytic result:
\begin{equation}
\label{eq:visibility.analytic}
V(\rho) = 2\pi \int_0^\infty I_\nu(r)\, J_0(2\pi \rho r)\, r\, \de r\,,
\end{equation}
where $\rho=\sqrt{u^2+v^2}$ is the deprojected \uv-baseline, $r$ is the angular distance from the centre and $J_0$ is the $0-$th order Bessel function of the first kind \citep{Pearson:1999aa}. For example, this approach has been recently used by \citep{Zhang:2016aa} to compare different brightness profiles to interferometric observations of protoplanetary discs.

Using Eq.~\eqref{eq:visibility.analytic} we compute analytical synthetic visibilities of four  brightness-profile templates with different features and we compare them to the visibilities output by the \func{sample*} functions.

\noindent
(a) a Gaussian disc with a Gaussian ring-like excess:
\begin{equation}
I_\nu(r) = \exp\left[-\left(\frac{r}{0.2\arcsec}\right)^2\right] + 0.3\,\exp\left[-\left(\frac{r-0.4\arcsec}{0.15\arcsec}\right)^2\right]\,,
\end{equation}
(b) a amooth Gaussian ring:
\begin{equation}
I_\nu(r) = \exp\left[-\left(\frac{r-0.5\arcsec}{0.1\arcsec}\right)^2\right]\,,
\end{equation}
(c) a sharp rectangular ring:
\begin{equation}
I_\nu(r) = \begin{cases}
1 & \mathrm{for}\ 0.2\arcsec \leq r \leq 0.5\arcsec\\
0 & \mathrm{otherwise}
\end{cases}
\end{equation}
(d) three Gaussian rings:
\begin{equation}
\begin{split}
I_\nu(r) = \exp\left[-\left(\frac{r-0.2\arcsec}{0.1\arcsec}\right)^2\right] + 0.7\,\exp\left[-\left(\frac{r-0.5\arcsec}{0.05\arcsec}\right)^2\right] + \\
+ 0.2\,\exp\left[-\left(\frac{r-0.7\arcsec}{0.03\arcsec}\right)^2\right]\,.
\end{split}
\end{equation}
The choice of these templates with smooth or sharp, small or large spatial features aims at reproducing typical brightness profiles that are used to fit real observations and also to check how well \galario samples the different spatial frequencies that characterize their visibility profiles.

\begin{figure*}
\centering
\resizebox{0.88\hsize}{!}{\includegraphics{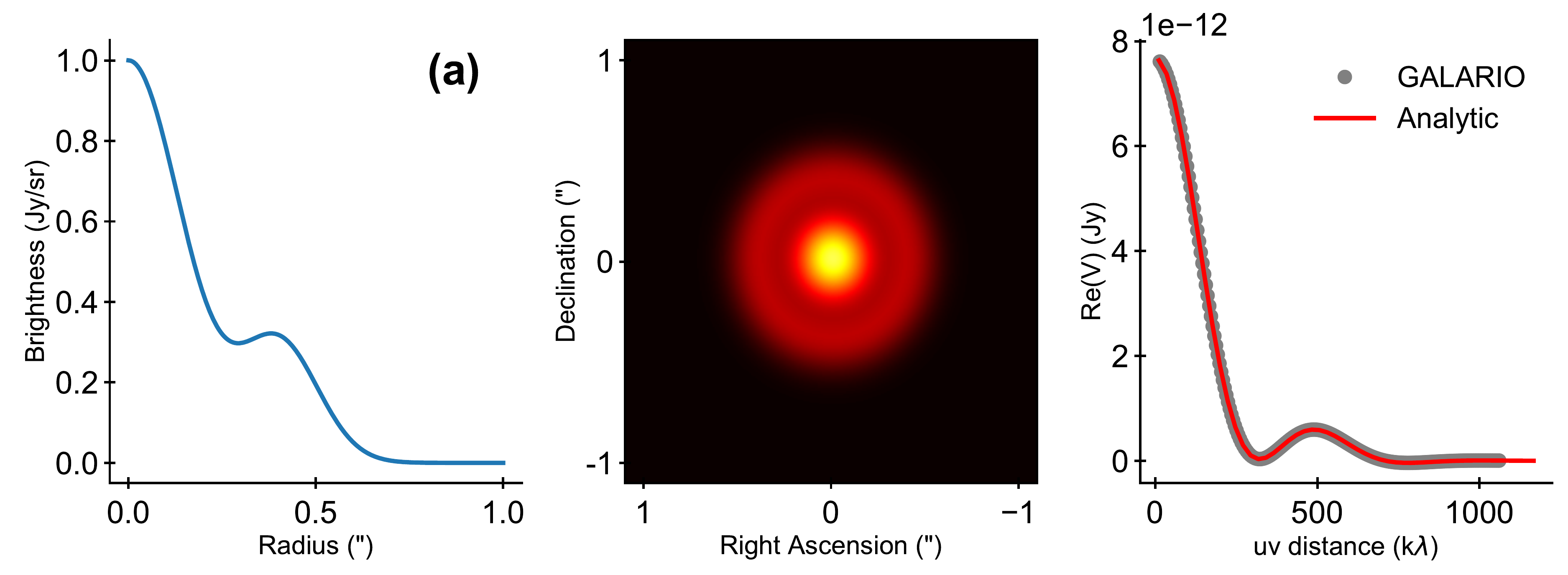}}
\resizebox{0.88\hsize}{!}{\includegraphics{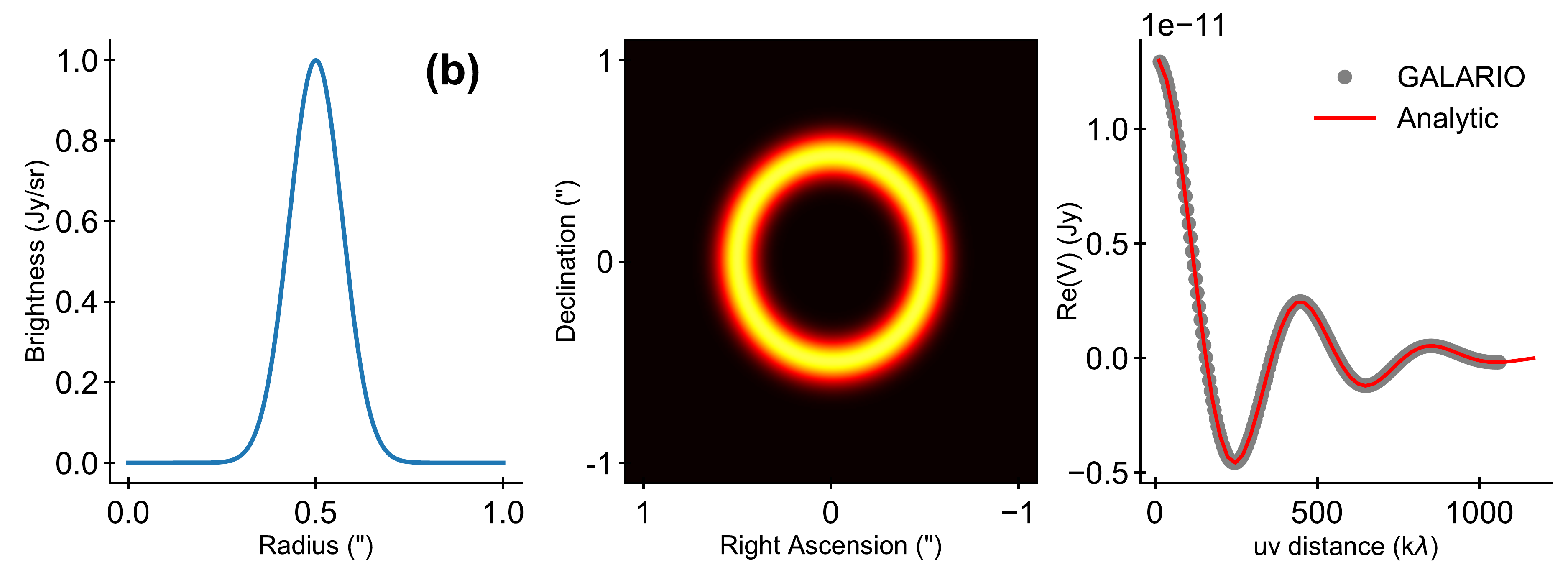}}
\resizebox{0.88\hsize}{!}{\includegraphics{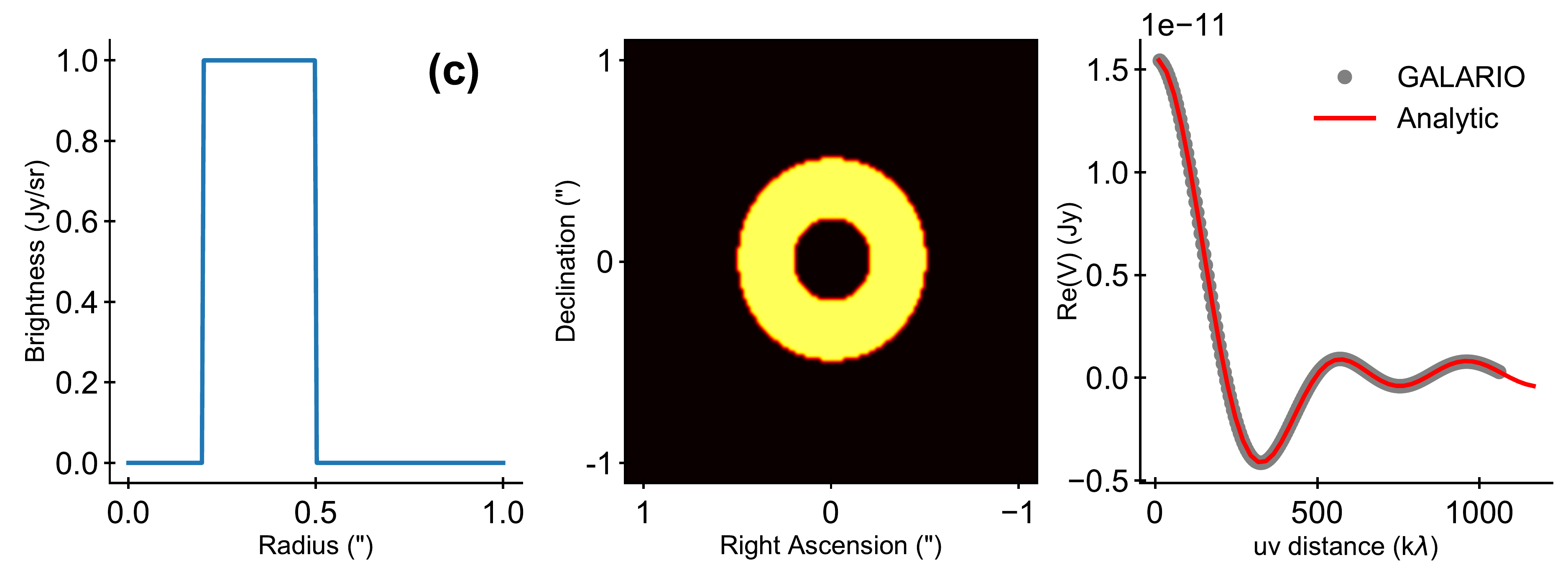}}
\resizebox{0.88\hsize}{!}{\includegraphics{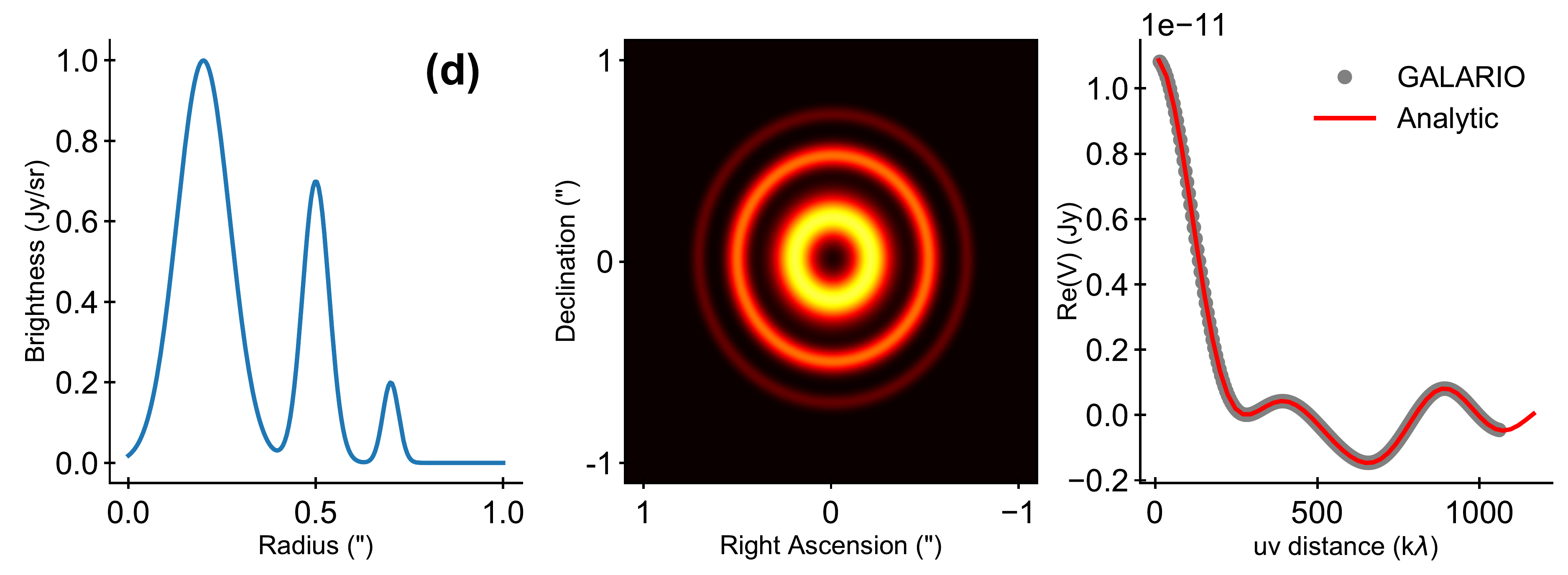}}
\caption{Results of the accuracy checks. For each of the templates we plot the radial brightness profile $I_\nu(r)$ (left panel), the image of the model (central panel), and the comparison of the synthetic visibilities (right panel).
The synthetic visibilities computed analytically (red lines) are compared to those computed by \galario (gray dots). 
The analytic synthetic visibilities have been sampled exactly in the same \uv locations of those computed by \galario but we show them as a continuous red line to aid the visual comparison.}
\label{fig:code.accuracy}
\end{figure*}

The visibilities are computed at \uv-points with baselines between 10\,k$\lambda$ and 1000\,k$\lambda$.
For reference, ALMA 1.3~mm observations in C43-5 configuration achieve a similar \uv coverage. The results presented below hold for different baseline extents and \uv-point locations; in Figure~\ref{fig:code.accuracy} we show just one of the several different configurations we tested.
For each of the templates we plot the radial brightness profile $I_\nu(r)$ (left panel), the image of the model (central panel), and the comparison of the synthetic visibilities (right panel). The \func{sampleProfile} and \func{sampleImage} functions yield identical synthetic visibilities at machine precision level, therefore in \figref{fig:code.accuracy} we just show the results for one of them, \func{sampleProfile}. Only the real part $\Re(V)$ of the synthetic visibilities is shown, since the imaginary part is identically zero for axisymmetric input images. 

In general we observe a very good agreement between the synthetic visibilities computed by \galario and and those computed analytically with Eq.~\eqref{eq:visibility.analytic}, as the deprojected visibility profiles in Fig~\ref{fig:code.accuracy} clearly show. \func{sampleProfile} and \func{sampleImage} model correctly the visibility profile of the templates at all the spatial frequencies. We performed quantitative checks on the discrepancy between the results and we find that the fractional difference between the sampled $\Re(V)$ values is generally smaller than $10^{-5}$. Only for a few data points where $\Re(V)$ is very close to zero does the fractional difference  reach a level of 0.1\%. We conducted numerous other consistency checks during the development of \galario that we do not report here -- e.g., comparing the output of the complex-to-complex Fourier transform with respect to the real-to-complex one, etc. -- but all are available as unit tests and can be executed from \galario's source code.

\section{Performance}
\label{sec:performance}

We now investigate the performance characteristics of \galario. All experiments
shown are performed on a desktop workstation with an Intel i7-6800K CPU with six
cores on one socket, hyperthreading, 3.4 GHz maximum frequency and 32 GB of RAM.
The machine also has an Nvidia GTX 1060 graphics card with 6 GB of RAM and 1280
CUDA cores. We also ran identical benchmarks on high-performance systems with 32
CPU cores and more powerful Nvidia P100 GPUs. While the exact timings differed,
we verified that our qualitative conclusions presented below also hold on these
much more expensive systems.

All results are for double precision only. We observed significant loss of
precision in the single-precision FFT for reasonably sized images beyond $512^2$
pixels that could affect scientific results whereas the double-precision FFT was
much more robust. Therefore we recommend double precision as the default mode in
\galario.

\subsection{Scaling with image size}

To justify the effort of creating this package, we consider an alternative
implementation of \func{chi2Profile} in standard \py without any explicit loops, using instead the widespread \comm{numpy} and
\comm{scipy} packages \citep{numpy} to do all the ``heavy lifting''. This represents a baseline solution that could be assembled in a short time without requiring deep thought. This \py version is shipped with \galario's unit tests {and is reported for completeness in Appendix~\ref{app:python.functions}.}

In \figref{fig:weak.scaling}
\begin{figure*}
\resizebox{\hsize}{!}{\includegraphics{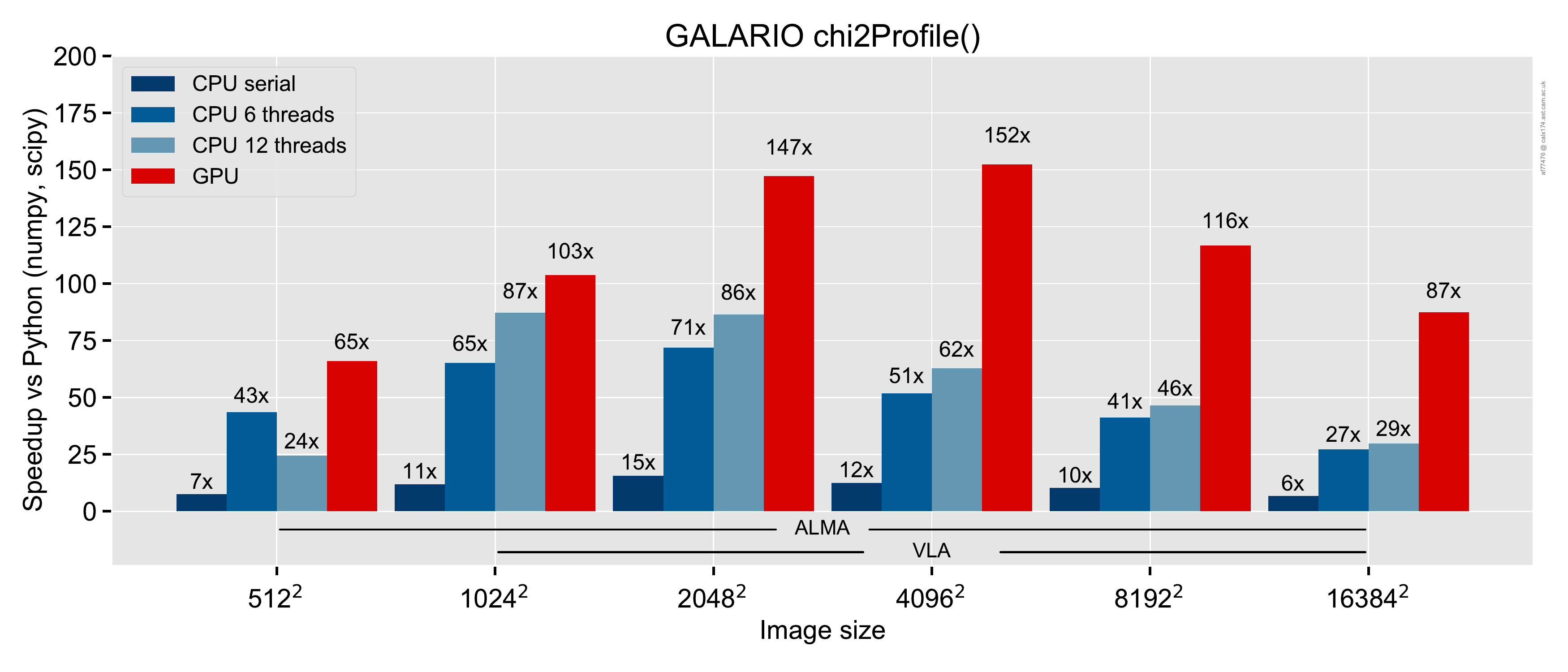}}
\caption{Scaling with image size: speed-up of the CPU and GPU version of \func{chi2Profile}. The CPU version is executed with 1, 6 and 12 threads. The speed-up is computed with respect to a \py version that relies on \comm{numpy} and \comm{scipy} and make no use of explicit loops. The absolute execution times are reported in \tbref{tb:timings.chi2Profile}.
The horizontal brackets highlight typical matrix sizes for realistic ALMA and {VLA} array configurations (cf. \tbref{tb:array.config.matrix.properties}). 
}
\label{fig:weak.scaling}
\end{figure*}
we show the scaling behaviour by calling
\galario's \func{chi2Profile} double-precision implementation for different
sizes of the input image varying from $512^2$ to $16384^2$ pixels. This is
repeated on the CPU with 1, 6, and 12 \openmp threads and on the GPU. The absolute timings are reported in \tbref{tb:timings.chi2Profile}, while \figref{fig:weak.scaling} presents the timings in terms of the speed-up relative to the \py-only baseline.
Contrary to common belief, even the serial CPU implementation of \galario is significantly faster than
the baseline, thus implying that there is a price to pay when relying on \comm{numpy} and
\comm{scipy} even though the relevant parts also execute compiled C code just
like \galario.
\begin{table}
\centering
\caption{Execution times of \func{chi2Profile}.}
\begin{tabular}{lrrrrr}
\toprule
&	&	\multicolumn{3}{c}{CPU}	&	GPU\\\cmidrule(lr){3-5}\cmidrule(lr){6-6}
$N_{lm}$	&	Python	&	Serial	&	6 threads	&	12 threads	& 	\\
(px)	&	(ms)	&	(ms)	&	(ms)	&	(ms)	&	(ms) \\
\midrule
512	&	  815	&	  109	&	   19	&	   33	&	   12	\\
1024	&	 1407	&	  120	&	   22	&	   16	&	   14	\\
2048	&	 2719	&	  175	&	   38	&	   31	&	   18	\\
4096	&	 5959	&	  478	&	  115	&	   95	&	   39	\\
8192	&	14702	&	 1440	&	  358	&	  317	&	  126	\\
16384	&	41895	&	 6204	&	 1536	&	 1411	&	  479	\\
\bottomrule
\end{tabular}
\label{tb:timings.chi2Profile}
\begin{flushleft}
{Notes}\quad Timings refer to the execution of the double-precision version of \func{chi2Profile} with
  $M=10^6$ visibility points.
\end{flushleft}
\end{table}

It turns out that because of the large amount of memory accesses inherent to our
algorithm, hyperthreading can help significantly. The biggest gain is $30 \%$ for
a $1024^2$ image comparing 6 to 12 threads; i.e. one to two threads per physical
core. In contrast, for the smallest image size $512^2$, the overhead of threads
leads to a performance penalty of about $50 \%$. It appears the optimum number
of threads has to be determined by trial and error.

Comparing the fastest CPU timing to the GPU timing, we observe that executing on
the GPU is about $30 \%$ to three times faster. A speed-up of two to five is
often observed when comparing optimized parallel implementations on CPU and GPU,
and bigger speed-ups occur when the baseline is serial or unoptimised CPU code
\citep{lee2010debunking}. The advantage of the GPU is the enormous number of threads
that can operate simultaneously, so it performs best if there are many
arithmetic operations per data unit. The disadvantage is that data transfer from
the CPU to the GPU and memory allocation on the GPU are much slower compared to
the CPU. The ideal application for the GPU then is for a large image that is
created on the GPU and need not be transferred from host memory.

\subsection{Strong scaling}

\begin{figure}
\resizebox{\hsize}{!}{\includegraphics{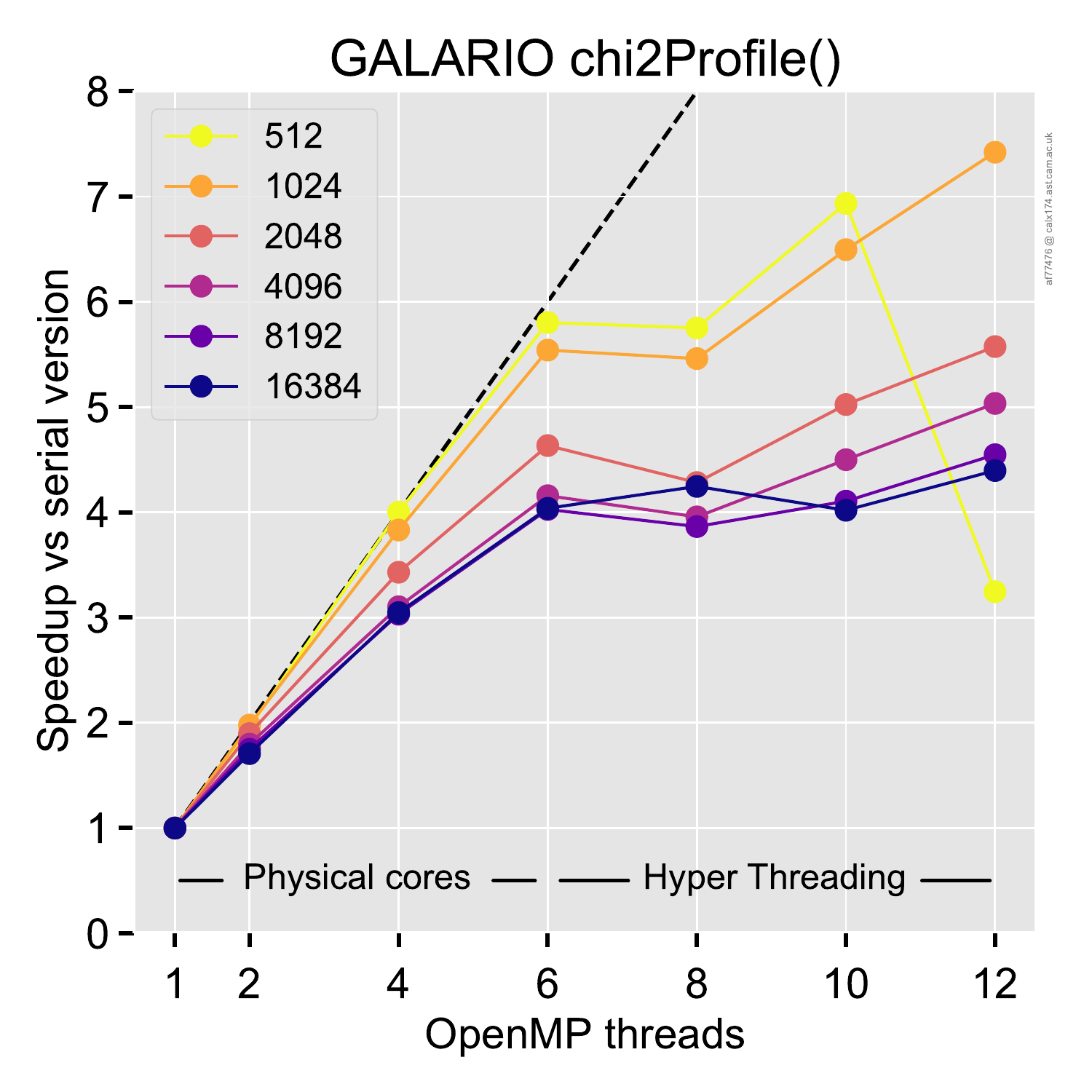}}
\caption{Strong scaling of the CPU version for different image sizes: speed-up (with respect to the serial version) for increasing computing power.}
\label{fig:strong.scaling.cpu}
\end{figure}

Taking the serial CPU code for one \openmp thread as the baseline,
\figref{fig:strong.scaling.cpu} shows the strong-scaling behaviour;
i.e., by how much the execution improves with more threads for a fixed
image size. We compute \func{chi2Profile} 300 times with identical
input parameters for each number of threads, and display the shortest
of the 300 times recorded.

For small images up to $1024^2$ pixels, the speed-up is nearly equal
to the number of threads until it reaches six, the number of physical
cores on our test machine. For larger images, the cache size becomes a
factor as threads compete for it and we have many memory-heavy
operations. The improvement up to six threads is still
monotonous. Using more threads than cores slightly hides the cache
misses, so for all image sizes except $512^2$, the highest performance
is attained for 12 threads, i.e. two threads per core, the maximum
supported by native hyperthreading.

\subsection{Profiling suboperations}

\begin{figure*}
\resizebox{0.49\hsize}{!}{\includegraphics{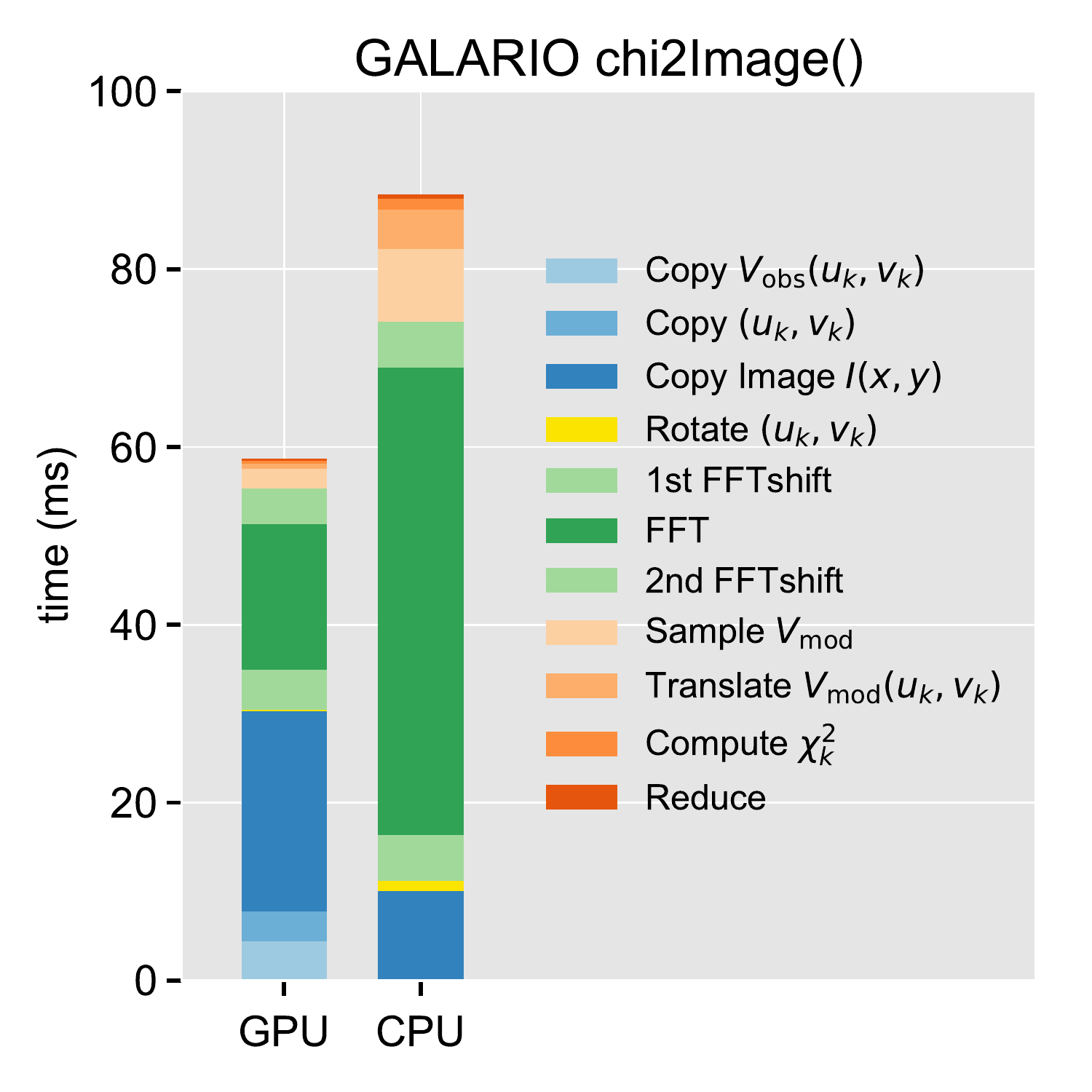}}
\resizebox{0.49\hsize}{!}{\includegraphics{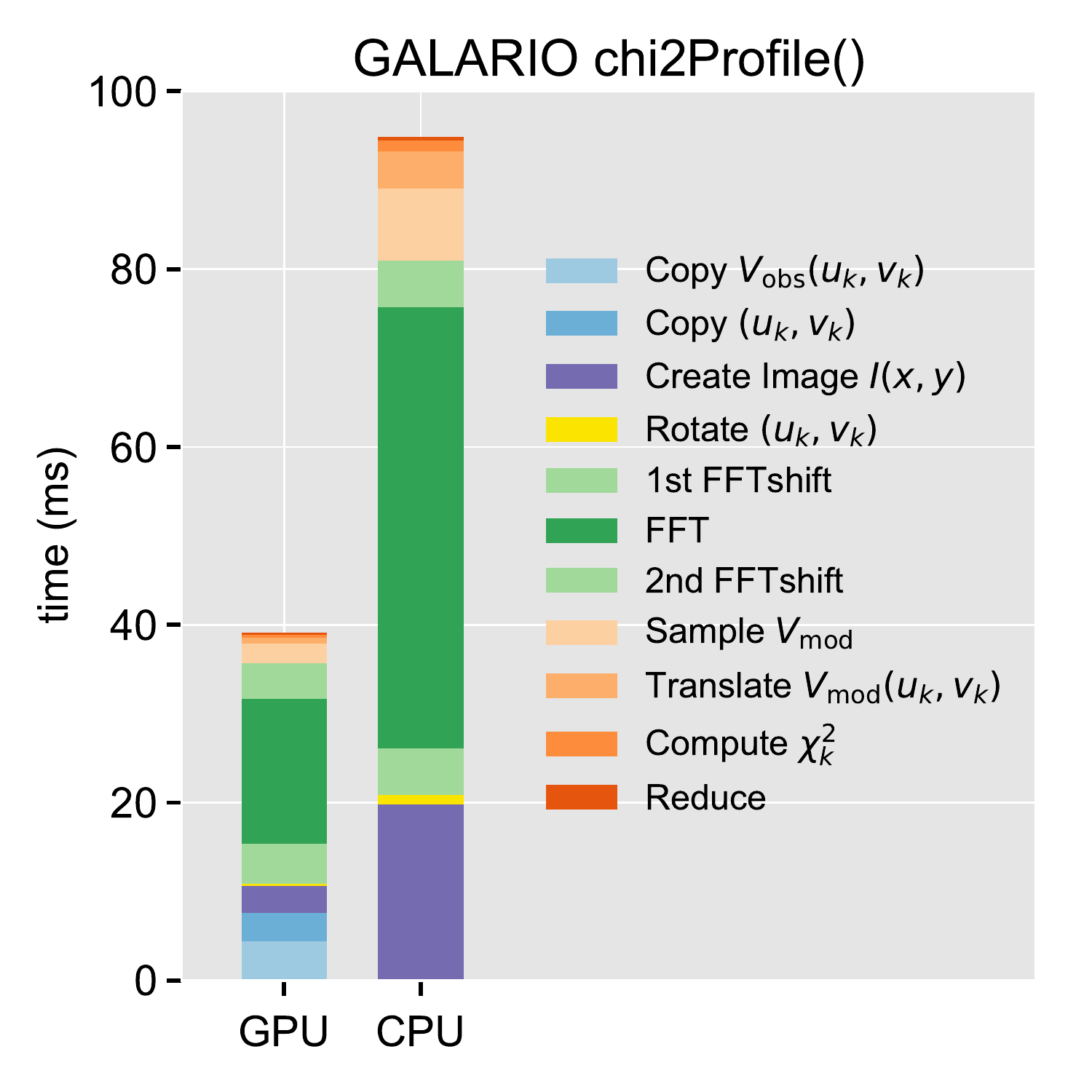}}
\caption{Execution times of the suboperations in \func{chi2Image} (left panel) and \func{chi2Profile} (right panel). Each vertical bar represents the smallest time across 20 calls of \func{chi2*} for a $4096^2$ image. The CPU version is run with the optimal configuration of threads for this image size. CPU to GPU (Host to Device) copy operations are coloured in tones of blue. On the GPU, \func{chi2Profile} dramatically reduces the time spent in copying with respect to \func{chi2Image} and, even accounting for the extra-time needed to create the image, it manages to be 1.5 times faster than \func{chi2Image}.}
\label{fig:profiling}
\end{figure*}

While \galario aims to be user friendly and accepts an input image
supplied by the user with \func{chi2Image}, it is generally
advantageous to create the image on the GPU. For the particular case
of an image created from a radial profile, \func{chi2Profile} only
transfers a radial profile equivalent to one row of pixels to the GPU,
creates the image on the GPU, then performs the same operations on the
image as \func{chi2Image}. In \figref{fig:profiling}, we show a
detailed break-down of the time that each of the suboperations
requires for both \func{chi2*} functions for a $4096^2$ image. We
compare only the optimal number of threads on the CPU (12 in this
case) to the GPU implementation. We repeated each function call 20
times and display the minimum time for each suboperation during the 20
runs.
Since no run features the minimum time for all suboperations,
the minimum time for \func{chi2*} across the 20 runs is slightly larger
than the sum of timings shown in \figref{fig:profiling}.

The important point of \figref{fig:profiling} is that all
compute-intensive operations are faster on the GPU compared to the
CPU, in particular the Fourier transform and the creation of the
image. But the data transfers partly reduce this advantage, either
because they are not needed at all on the CPU (for example the copy of
read-only data like observations) or because the PCI express bus has a
much reduced bandwidth and higher latency compared to accesses to the
main CPU memory.

For a large image, copying to the device actually takes longer than
the FFT, that is why \func{chi2Image} is nearly \SI{20}{ms}, or
\SI{50}{\%}, slower than \func{chi2Profile} on the GPU, and that does
not even account for the time needed to create the image on the CPU,
which in this example takes another \SI{20}{ms}.  On the CPU,
\func{chi2Profile} is only about \SI{15}{\%} faster than
\func{chi2Image}.

\section{Conclusions}
\label{sec:conclusions}
In this paper we have presented \galario, a GPU accelerated library for analysing radio interferometer observations. Distributed under the open source GNU LGPLv3, \galario is actively developed at \githubrepo and can be easily installed on machines with different configurations.

Unlike single dishes, which directly measure the brightness of a source over a continuous region of the sky, radio interferometers measure its Fourier transform, sampling it at discrete locations and producing a collection of complex visibilities.
The computational effort required to compare a model prediction to an observational data set of complex visibilities has increased dramatically in the last years due to the improved angular resolution and data rate of modern radio interferometers, which require larger matrix sizes and involve hundreds of thousands of visibility points.

The process of computing synthetic visibilities from a model brightness involves several time-consuming matrix operations such as Fourier transforms, quadrant swaps, and interpolations. These operations have to be performed once for every likelihood evaluation, and the likelihood is called thousands or even a million times in the normal workflow required to fit a model to the data.

In this context, \galario leverages the computing power of modern Graphical processing units (GPUs) to accelerate the computation of the synthetic visibilities, thus reducing the overall execution time of the likelihood. For ease of use, \galario offers dedicated functions that produce directly the weighted $\chi^2$ of the model for the given observations. Such functions can be easily included in any analysis scheme, be it a Markov Chain Monte Carlo sampler or a classical $\chi^2$ optimizer.
Moreover, thanks to its modularity, \galario can be used to fit simultaneously several observations at different wavelengths, thus speeding up even the most demanding multi-wavelength analyses.

\galario is easy to use -- computing the synthetic visibilities from a model image can be done in one line of code -- and easy to adopt in existing code -- Python wrappers to the underlying \CC and \CUDA code are available for all the functions. The design of \galario, with symmetric CPU and GPU versions of all the functions, allows the user to develop highly reusable code that can be executed both on CPUs and on GPUs with minimal changes, ensuring considerable speed-ups also on machines without a GPU. 

In terms of performances, \galario is faster than a standard Python implementation of the same functionalities by up to 150 times on the GPU and up to 90 times on the CPU. We note that these speed-ups are achievable not only on top-tier GPUs, but also on affordable desktop-class ones.

In the future releases of \galario we plan to generalise the implementation of the synthesis imaging equations by including the primary beam correction and a proper treatment of non-coplanar baselines (relevant for wide field imaging). Moreover, several new features will be added, including the multi-wavelength synthesis of brightness models with spectral dependence.
\section*{Acknowledgements}
{ 
The authors thank the anonymous referee for a constructive report that helped improving the clarity of the paper in many points. MT is grateful to Padelis P. Papadopoulos, Attila Juhasz, Luca Matr\`a and Sebastian Marino for the numerous insightful discussions on radio interferometry.}
MT has been supported by the DISCSIM project, grant agreement 341137
funded by the European Research Council under ERC-2013-ADG.
FB and LT acknowledge support by the DFG cluster of excellence
Origin and Structure of the Universe (\href{http://www.universe-cluster.de}{www.universe-cluster.de}).
The initial development of \galario was boosted by the GPU Hackathon at the TU Dresden in February 2016 thanks to two supportive mentors, Thorsten Hater and Andreas Herten. 
{The GPU Hackathon is a collaboration between and used resources of both TU Dresden and the Oak Ridge Leadership Computing Facility at the Oak Ridge National Laboratory. Oak Ridge National Laboratory  is supported by the Office of Science of the U.S. Department of Energy under Contract No. DE-AC05-00OR22725.}
Some development and testing of the GPU implementation has taken place on GPU machines by the Leibniz Supercomputing Center's data lab. Testing has also been carried out on the Hydra computing cluster by the Max Planck Gesellschaft, with support by Paola Caselli. 
\galario 1.0 has the following Zenodo reference \href{https://doi.org/10.5281/zenodo.889991}{https://doi.org/10.5281/zenodo.889991}.



\bibliographystyle{mnras}
\bibliography{mt_disks,more_refs} 

\appendix
\section{Installation}
\label{app:install}
\galario is actively developed online at \githubrepo. Contributions are welcome and we invite users that encounter problems in using \galario to report them at \githubissues.

{ 
The easiest way to install \galario is via \comm{conda}, the package manager of the Anaconda Python distribution\footnote{link}, which ensures all the dependencies are installed automatically. With conda, the user gets access to \galario C/C++ and Python bindings, both with support for multi-threading. The installation command is as easy as
\begin{bashcode}
conda install -c conda-forge galario
\end{bashcode}
Due to technical limitations, the \comm{conda} package does not support GPUs at the time of writing. In order to use the GPU version, \galario must be compiled by hand as follows. First,}
download the latest stable version from the repository with
\begin{bashcode}
git clone https://github.com/mtazzari/galario.git
\end{bashcode}
\galario works with both \py 2 and 3, and to simplify the build we suggest to work in a \py virtual environment.  Instructions on how to create an environment are reported in the online documentation.

Once downloaded, \galario can be installed with:
\begin{bashcode}
cd galario
mkdir build && cd build
cmake .. && make
\end{bashcode}
which will compile the CPU version of \galario and, if a GPU is present on your machine, also the GPU version.
The \comm{cmake} command takes care of adapting the compilation instructions to the compilers and the libraries available on your machine.

Once compiled, \galario can be installed with
\begin{bashcode}
sudo make install
\end{bashcode}
or, in the case the user has no root privileges, an installation path can be specified with
\begin{bashcode}
cmake -DCMAKE_INSTALL_PREFIX=/path/to/galario/ ..
make install
\end{bashcode}
This installs the \CC libraries of \galario in \comm{path/to/galario/lib} and the \py libraries in the currently active Python environment.

A full list of system requirements and detailed instructions to compile \galario on different systems are available in the online documentation at \githubdocs.

\section{Performance (continued)}
\label{app:performance}

\begin{figure*}
\resizebox{\hsize}{!}{\includegraphics{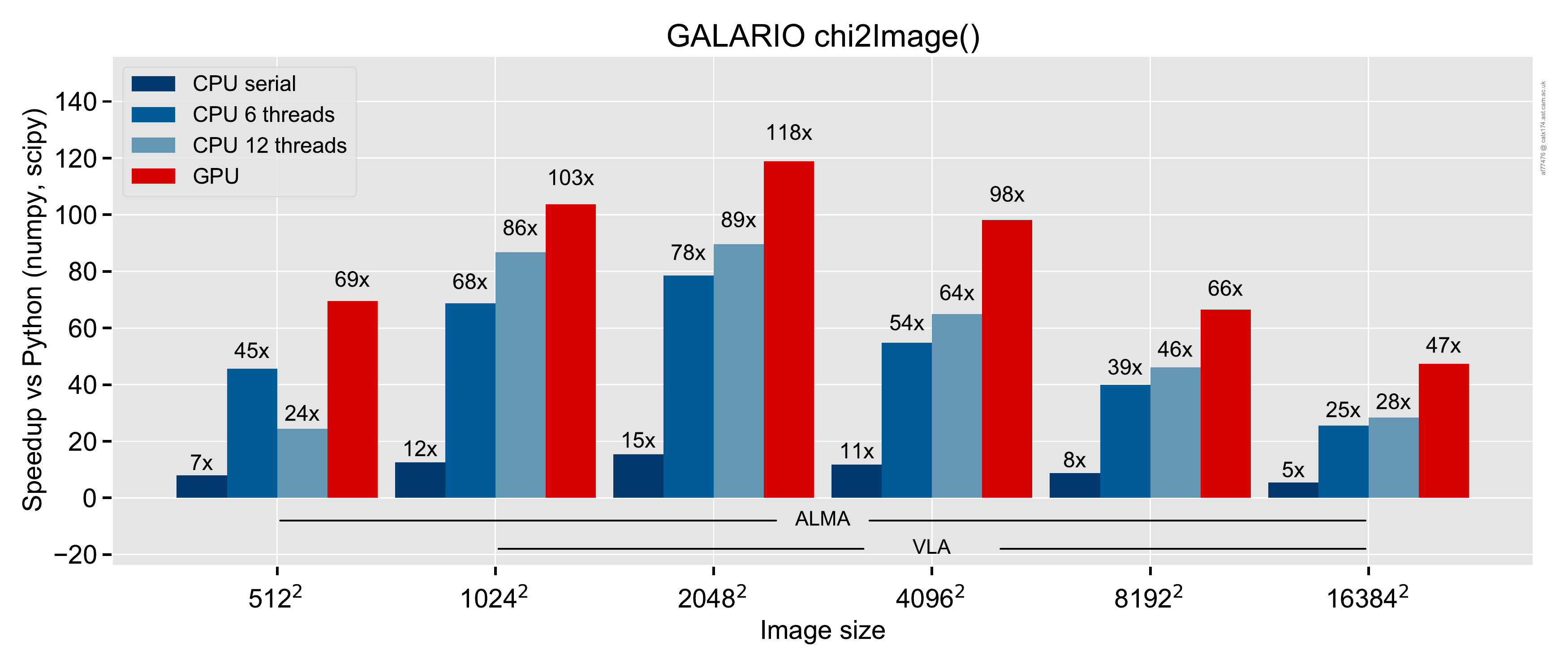}}
\caption{Scaling with image size: speed-up of the CPU and GPU version of \func{chi2Image}. The CPU version is executed with 1, 6 and 12 threads. The speed-up is computed with respect to a \py version that relies on \comm{numpy} and \comm{scipy} and make no use of explicit loops. The absolute execution times are reported in \tbref{tb:timings.chi2Image}.
The horizontal brackets highlight typical matrix sizes for realistic ALMA and {VLA} array configurations (cf. \tbref{tb:array.config.matrix.properties}).}
\label{fig:weak.scaling.chi2Image}
\end{figure*}

\begin{figure}
\resizebox{\hsize}{!}{\includegraphics{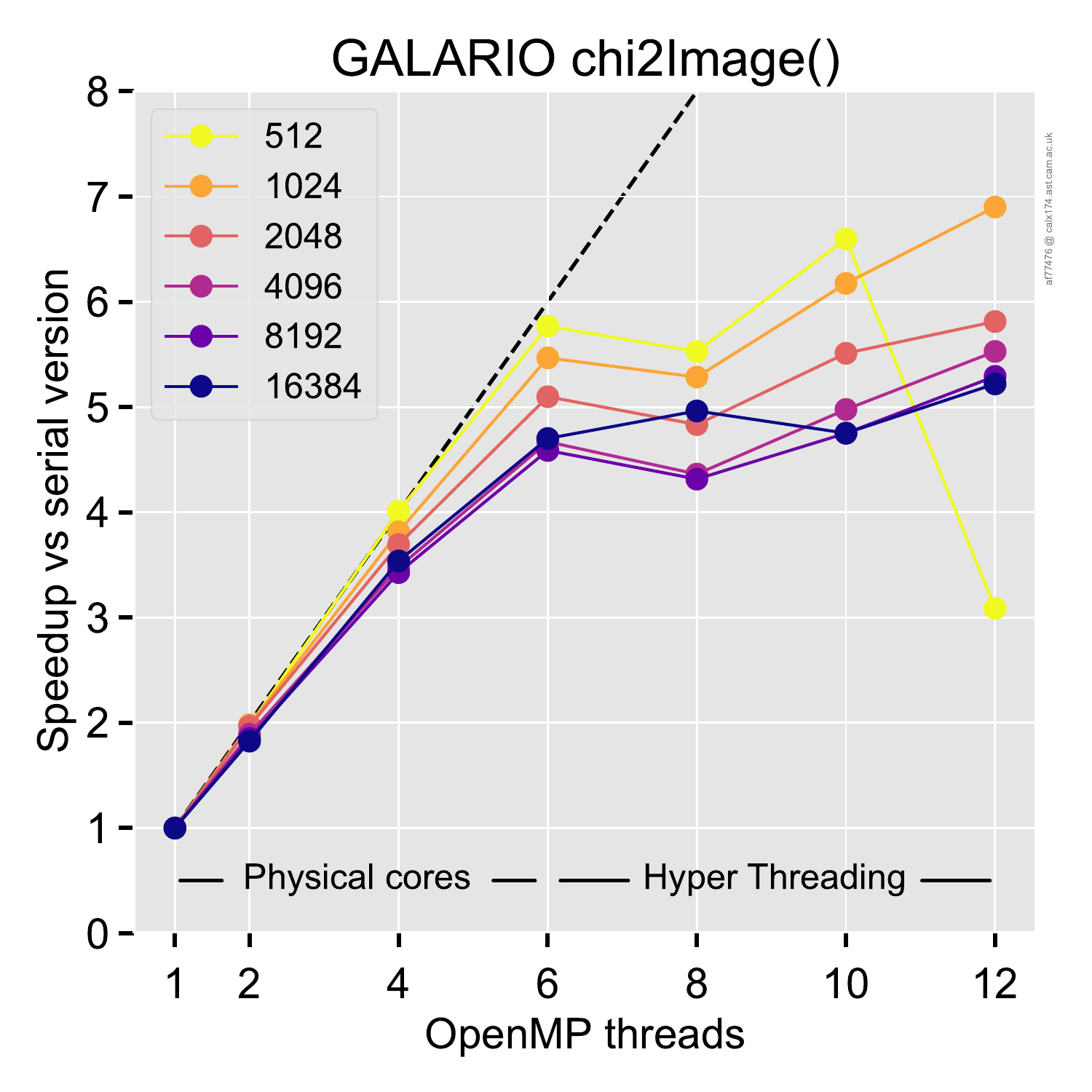}}
\caption{Strong scaling of the CPU version for different image sizes: speed-up (with respect to the serial version) for increasing computing power.}
\label{fig:strong.scaling.cpu.chi2Image}
\end{figure}

In Sect.~\ref{sec:performance} we presented the performance of \func{sampleProfile}. For completeness, in this Appendix we report analogous performance measurements conducted for \func{sampleImage}. \figref{fig:weak.scaling.chi2Image} shows the scaling of the CPU and GPU version as a function of matrix size, while \figref{fig:strong.scaling.cpu.chi2Image} shows the  scaling of the CPU version for increasing computing power. The absolute time measurements are reported in \tbref{tb:timings.chi2Image}.

\begin{table}
\centering
\caption{Execution times of \func{chi2Image}}
\begin{tabular}{lrrrrr}
\toprule
&	&	\multicolumn{3}{c}{CPU}	&	GPU\\\cmidrule(lr){3-5}\cmidrule(lr){6-6}
$N_{lm}$	&	Python	&	Serial	&	6 threads	&	12 threads	& 	\\
(px)	&	(ms)	&	(ms)	&	(ms)	&	(ms)	&	(ms) \\
\midrule
512	&	  799	&	  101	&	   18	&	   33	&	   11	\\
1024	&	 1385	&	  110	&	   20	&	   16	&	   13	\\
2048	&	 2651	&	  172	&	   34	&	   30	&	   22	\\
4096	&	 5747	&	  489	&	  105	&	   88	&	   59	\\
8192	&	13829	&	 1589	&	  346	&	  300	&	  208	\\
16384	&	38311	&	 7038	&	 1497	&	 1348	&	  810	\\
\bottomrule
\end{tabular}
\begin{flushleft}
{Notes}\quad Timings refer to the execution of the double-precision version of \func{chi2Image} with $M=10^6$  visibility points.
\end{flushleft}
\label{tb:timings.chi2Image}
\end{table}

{
\section{Accuracy (continued)}
\label{app:accuracy}
In this Section we present some of the results of the accuracy checks that we carried out against the NRAO CASA package. We ran a large suite of tests, here we show only some representative cases for different model images. 

In all these tests we used the results of the \func{sampleImage} function of \galario and of the \comm{ft} command of CASA, which are designed to perform the same operation: compute the sampled visibilities $V(u_k, v_k)$ for a given model image. For each image of the source brightness $I_\nu(l,m)$, we computed the visibilities in two ways: (i) we applied \galario's \func{sampleImage} to the 2d matrix containing the image; (ii) we exported the image to a FITS file, imported it in CASA with the \comm{importfits} task and then Fourier-sampled it with the \comm{ft} task (invoked with \comm{usescratch} option set to \comm{True}).

Figure~\ref{fig:app.accuracy} shows the comparison for three different input models, reporting a central cut of the image $I_\nu(l,m)$ (left column), a comparison of the amplitude $[Re(V)^2+Im(V)^2]^{1/2}$ (central column) and a comparison of the phase $\arctan[Im(V)/Re(V)]$ (right column). In the amplitude and phase plots, the bottom panels represent, respectively, the relative and absolute difference between the \galario and the CASA results (except in the first row, where the results are each compared to the analytic solution). The \uv-points used for the comparison represent a realistic \uv coverage for an ALMA observation, with baselines in the range 11-1370k$\lambda$, corresponding to a maximum recoverable scale $\theta_\mathrm{MRS}\sim 10\arcsec$ and an angular resolution $\theta_\mathrm{res}\sim 0.15\arcsec$.

In the first row the image is an axisymmetric model centred at the phase centre. By definition its visibility function is real-valued and has an analytic solution given by the Hankel transform in Eq.~\eqref{eq:visibility.analytic}. We therefore compare the results given by \galario and by CASA against the analytic solution. Both \galario and CASA reproduce very well the analytic amplitude within 0.1\% up to approximately 750\,k$\lambda$. \galario is slightly more accurate than CASA at longer baselines, up to 875\,k$\lambda$. The frequent spikes in the relative difference are due to the very sharp shape of the lobes, and occur only at the amplitude minima. Since the model is axisymmetric, the phase should be identically zero at every baseline: the phase plot clearly shows that \galario is almost 10 orders of magnitude more accurate than CASA in reproducing the null phase.

In the second row the image is made by multiple sources displaced across the field of view, and in the third row by a mock observation of Saturn (not to scale). These images have been chosen because they exhibit structures spanning a wide range of spatial scales, and therefore are useful to probe the accuracy of the codes across a wide range of spatial frequencies. In both cases, the amplitude and phase comparison shows a good agreement between the results obtained by \galario and CASA.  In the second row the vast majority of \uv-points agree better than 0.5-1\%, while in the third row the agreement is slightly worse, within 1-4\%. The discrepancies are significant only for the visibility phase in a handful of \uv-points (10-100) out of a total of approximately $10^5$ visibilities. Additionally, we compared the synthesized images (not reported here) produced from the visibilities computed by CASA and \galario and we find that they also are in very good agreement. We note that the slightly larger discrepancies found in the third row might be due to the fact that the CASA \comm{ft} task is likely applying a correction for wide field effects that is not included in the first release of \galario used here: this might be marginally relevant for the source brightness used in the third row, which is more extended than those used in the first two rows. However, due to the lack of documentation in the CASA package, it was unclear how to disable such correction for wide field effects when using the \comm{ft} task.
 
\begin{figure*}
\resizebox{0.32\hsize}{!}{\includegraphics{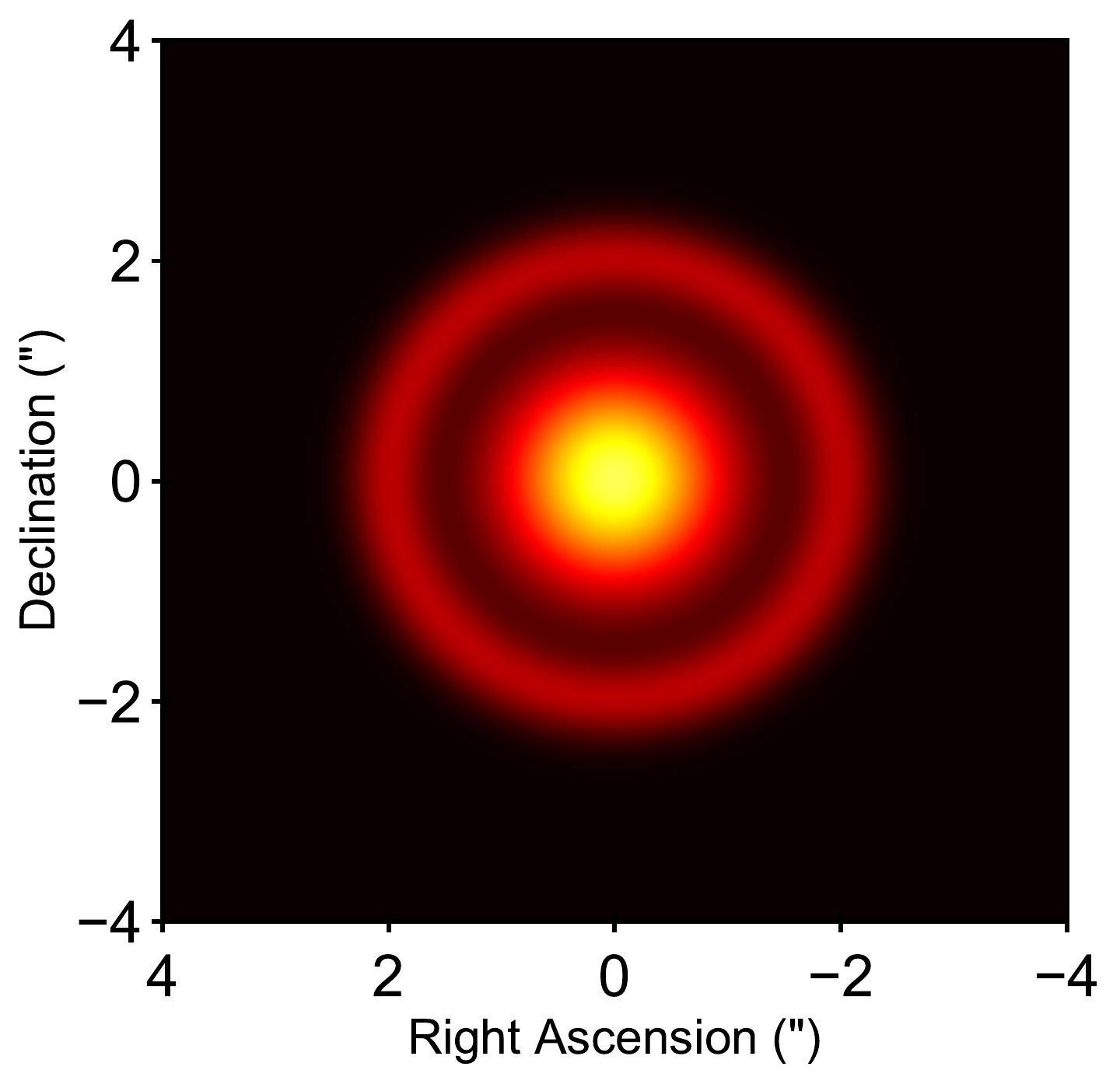}}
\resizebox{0.32\hsize}{!}{\includegraphics{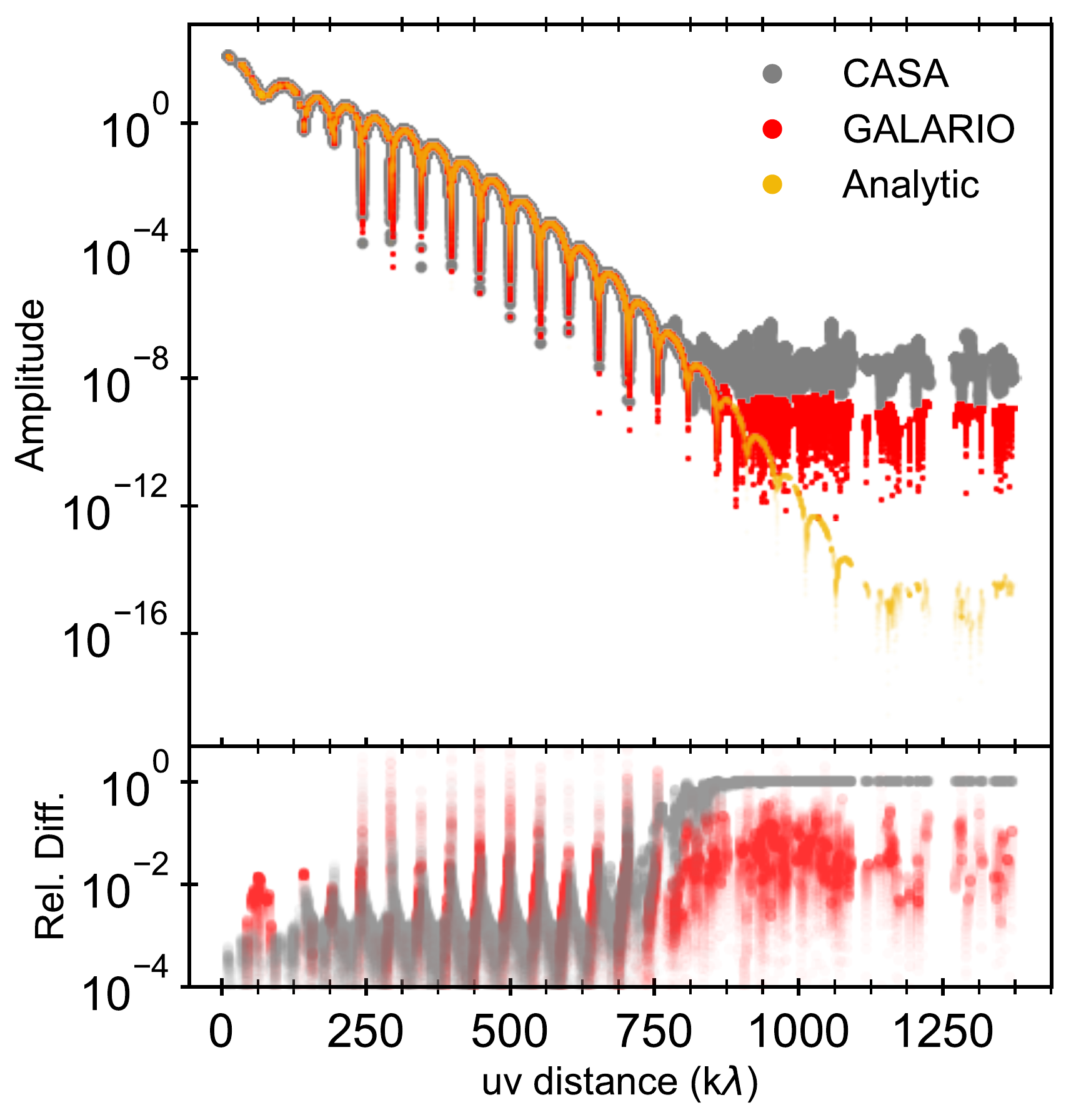}}
\resizebox{0.32\hsize}{!}{\includegraphics{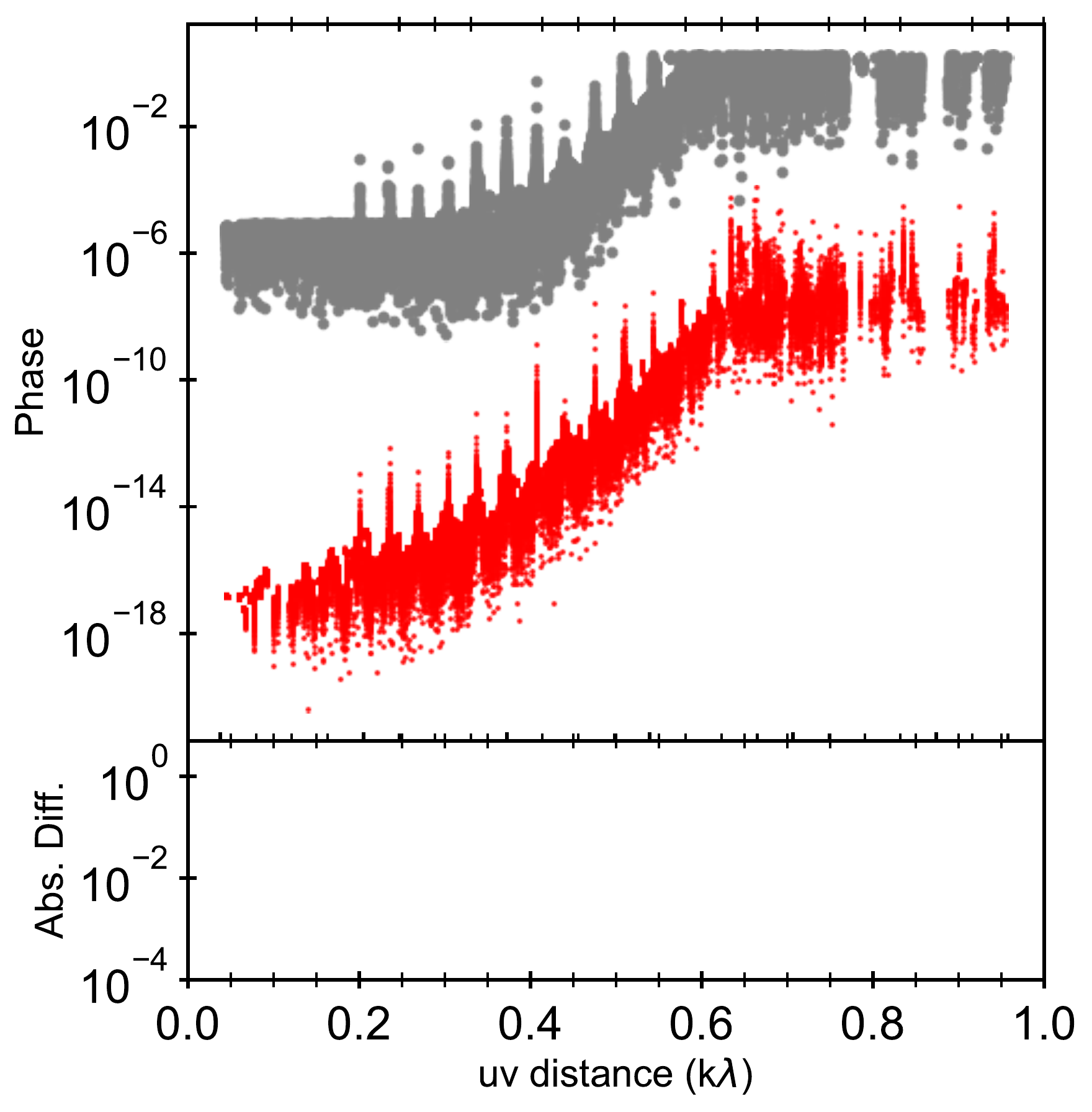}}\\
\resizebox{0.32\hsize}{!}{\includegraphics{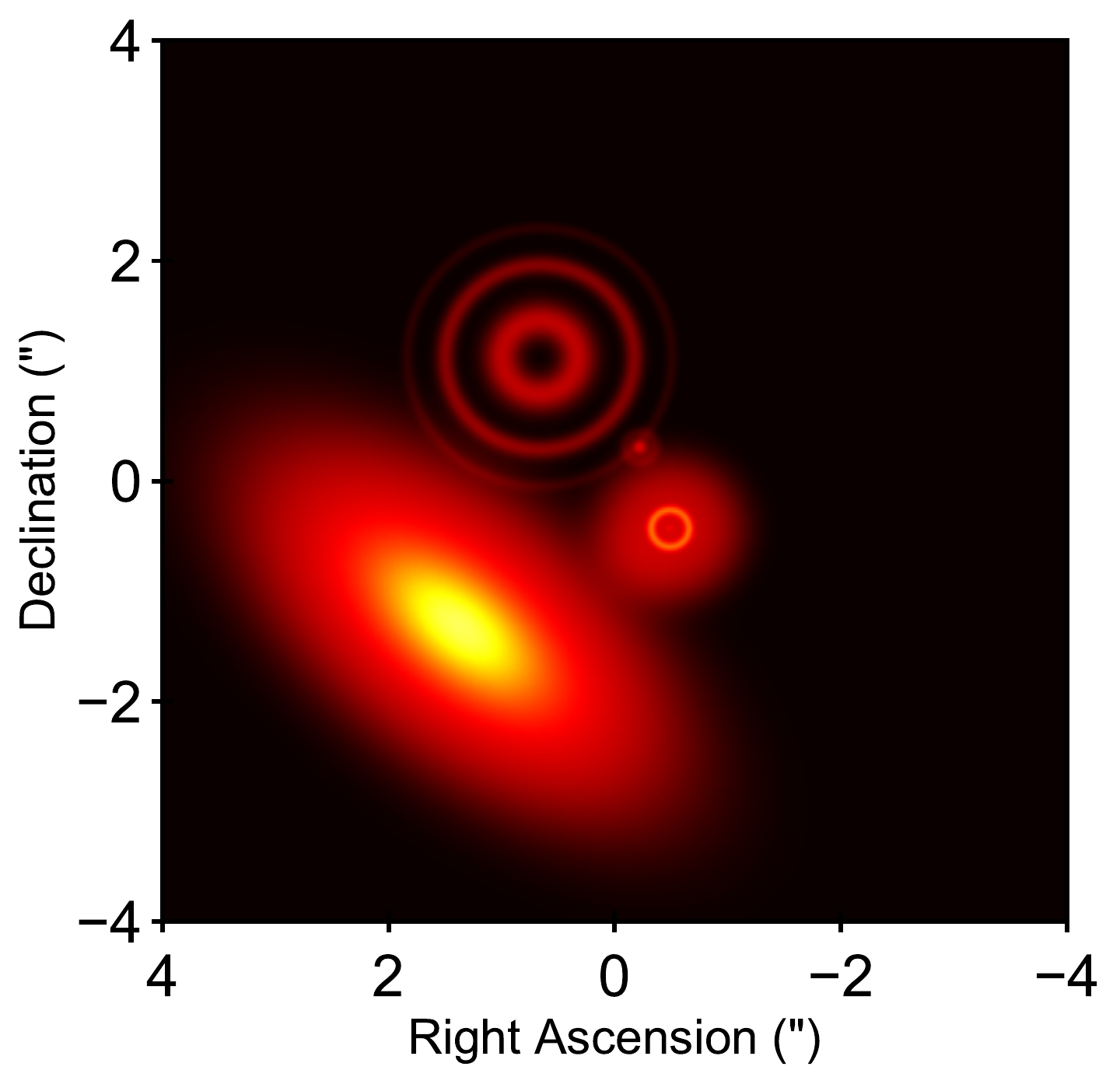}}
\resizebox{0.32\hsize}{!}{\includegraphics{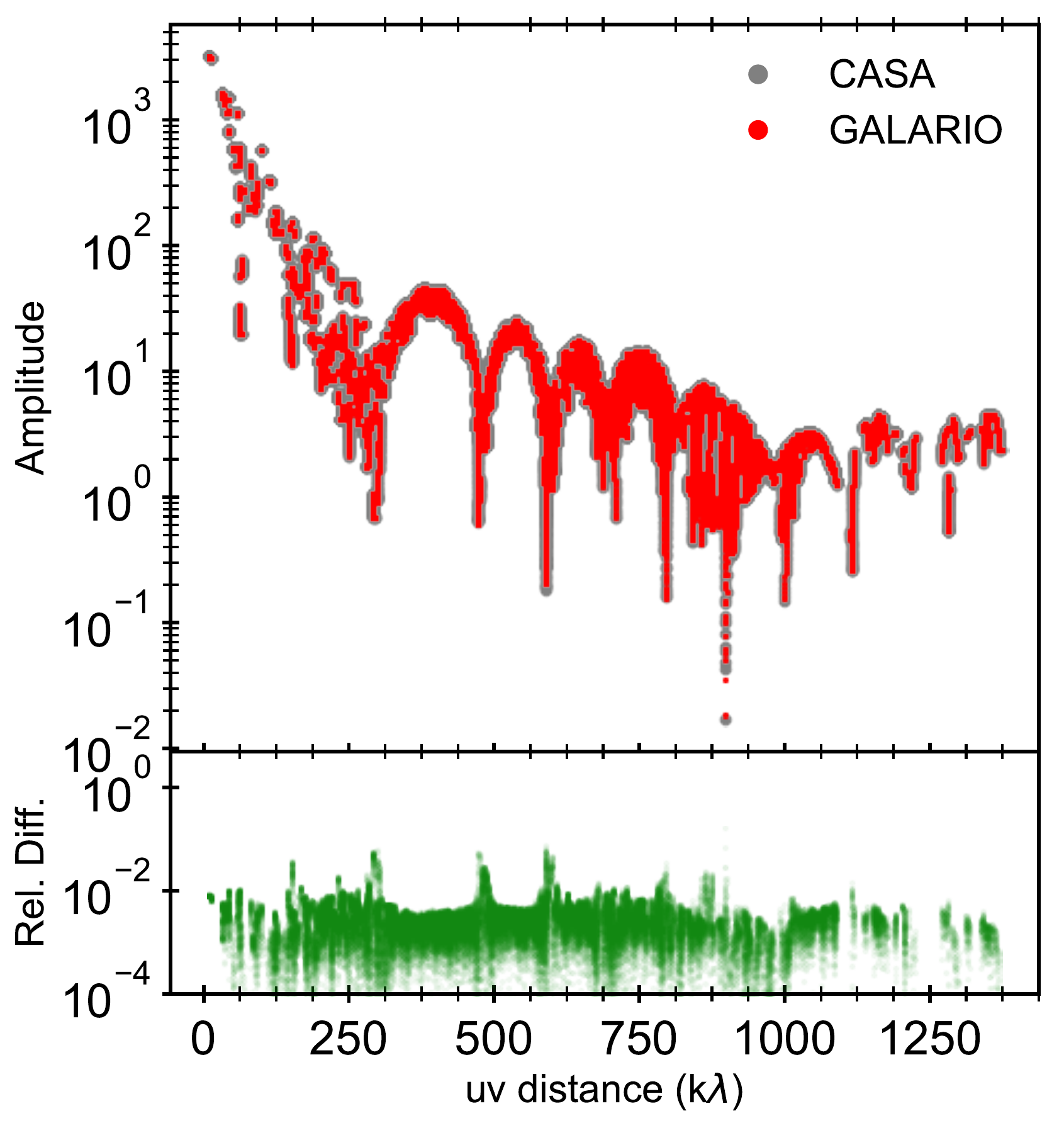}}
\resizebox{0.32\hsize}{!}{\includegraphics{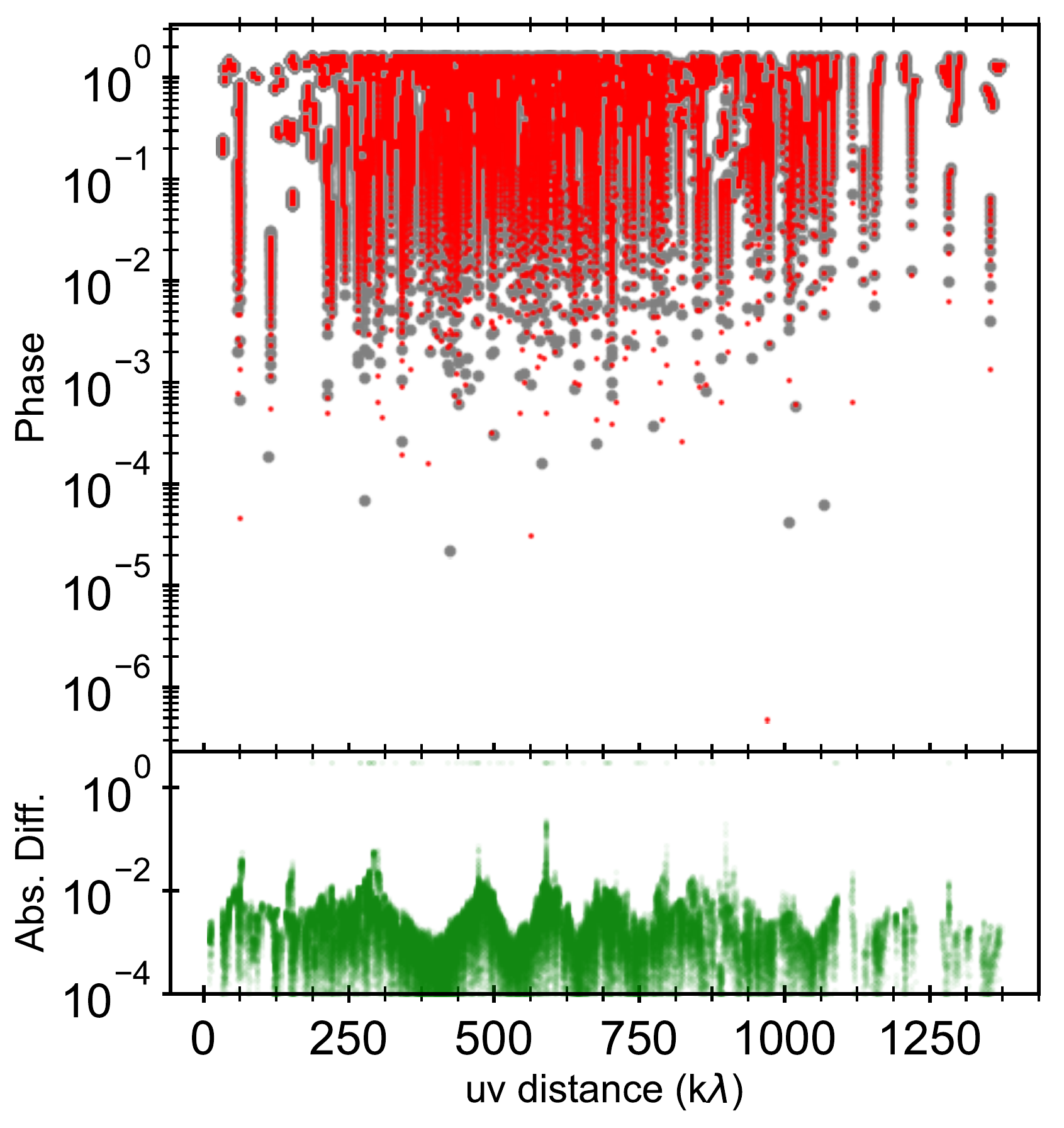}}\\
\resizebox{0.32\hsize}{!}{\includegraphics{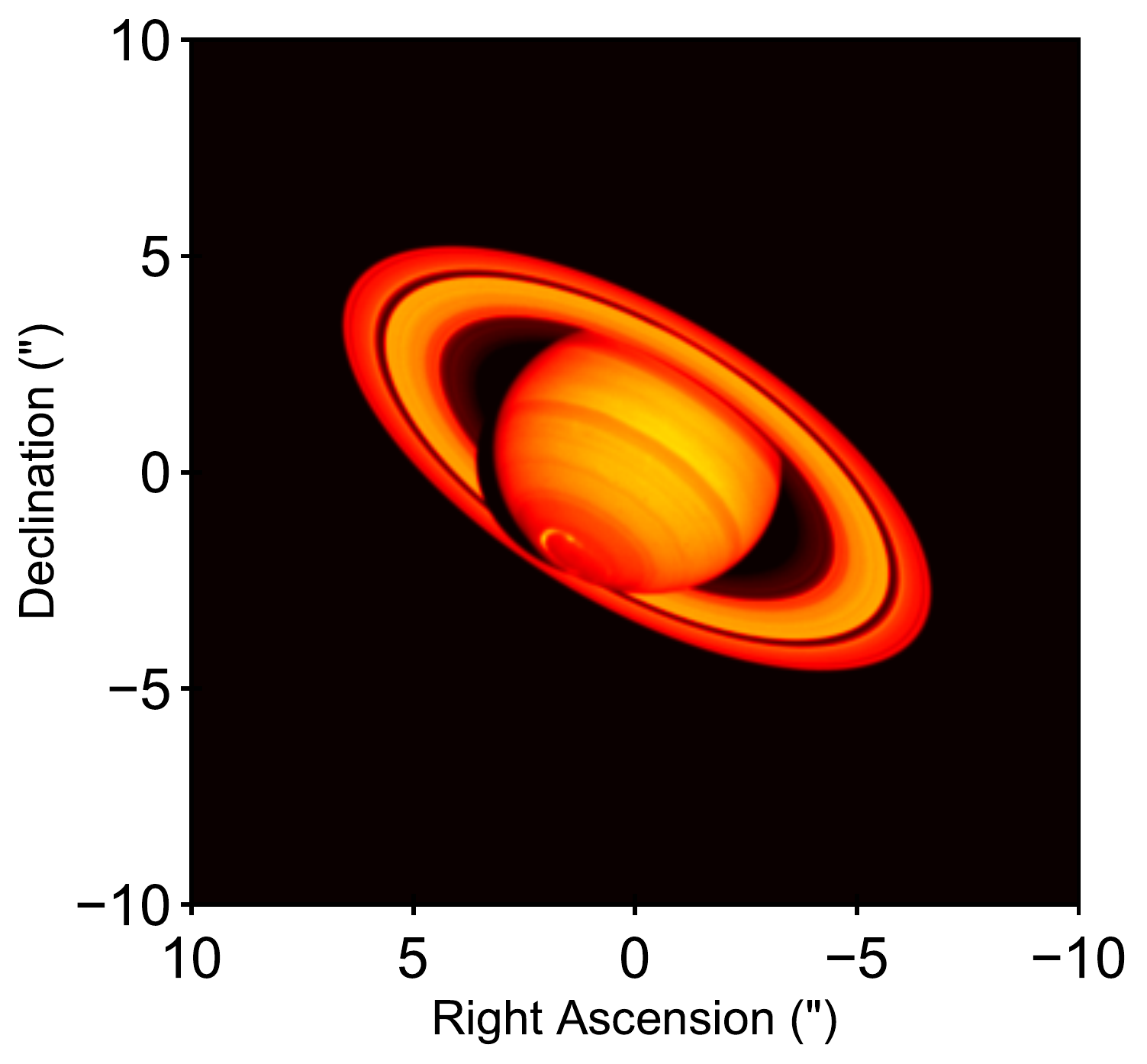}}
\resizebox{0.32\hsize}{!}{\includegraphics{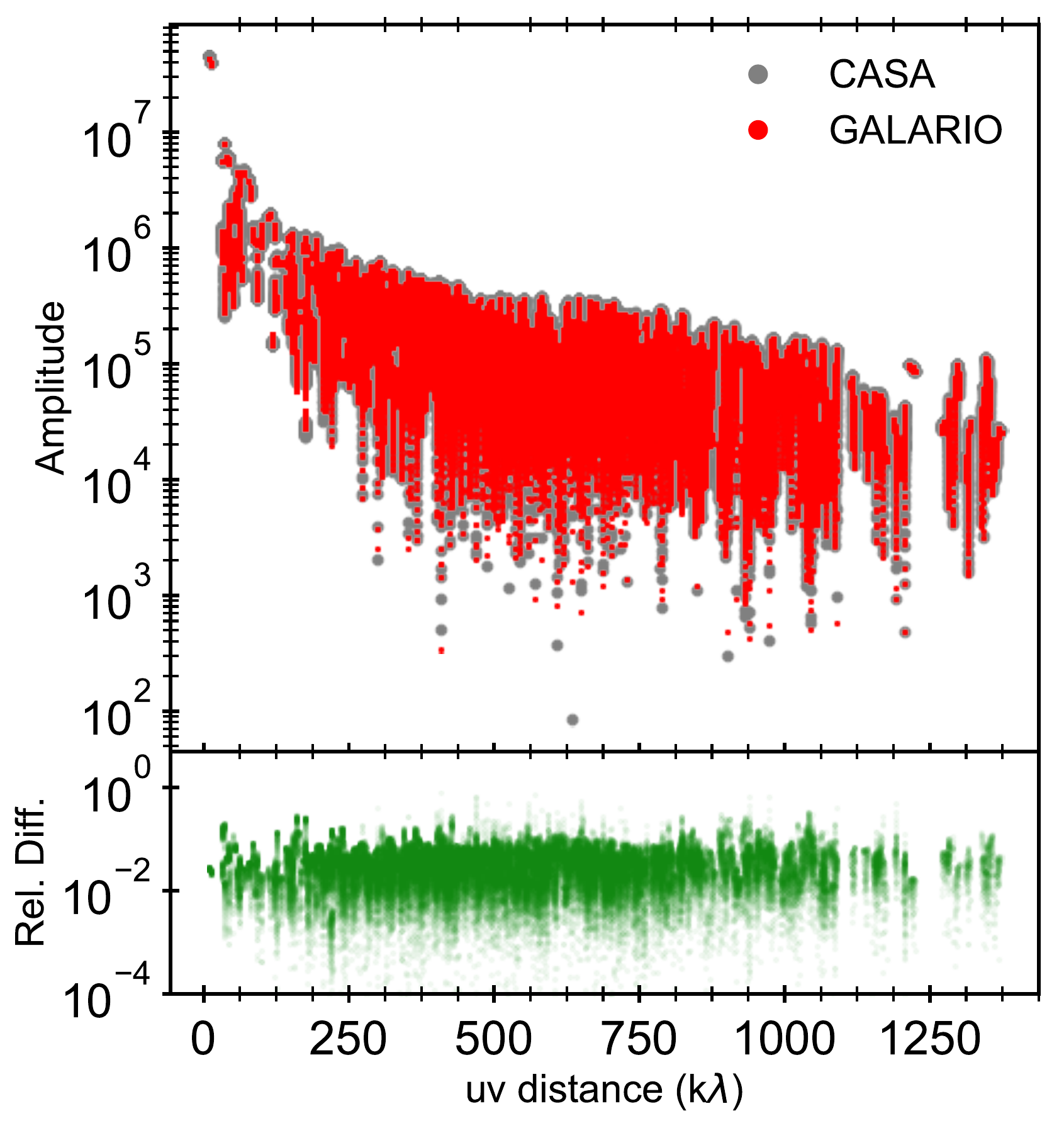}}
\resizebox{0.32\hsize}{!}{\includegraphics{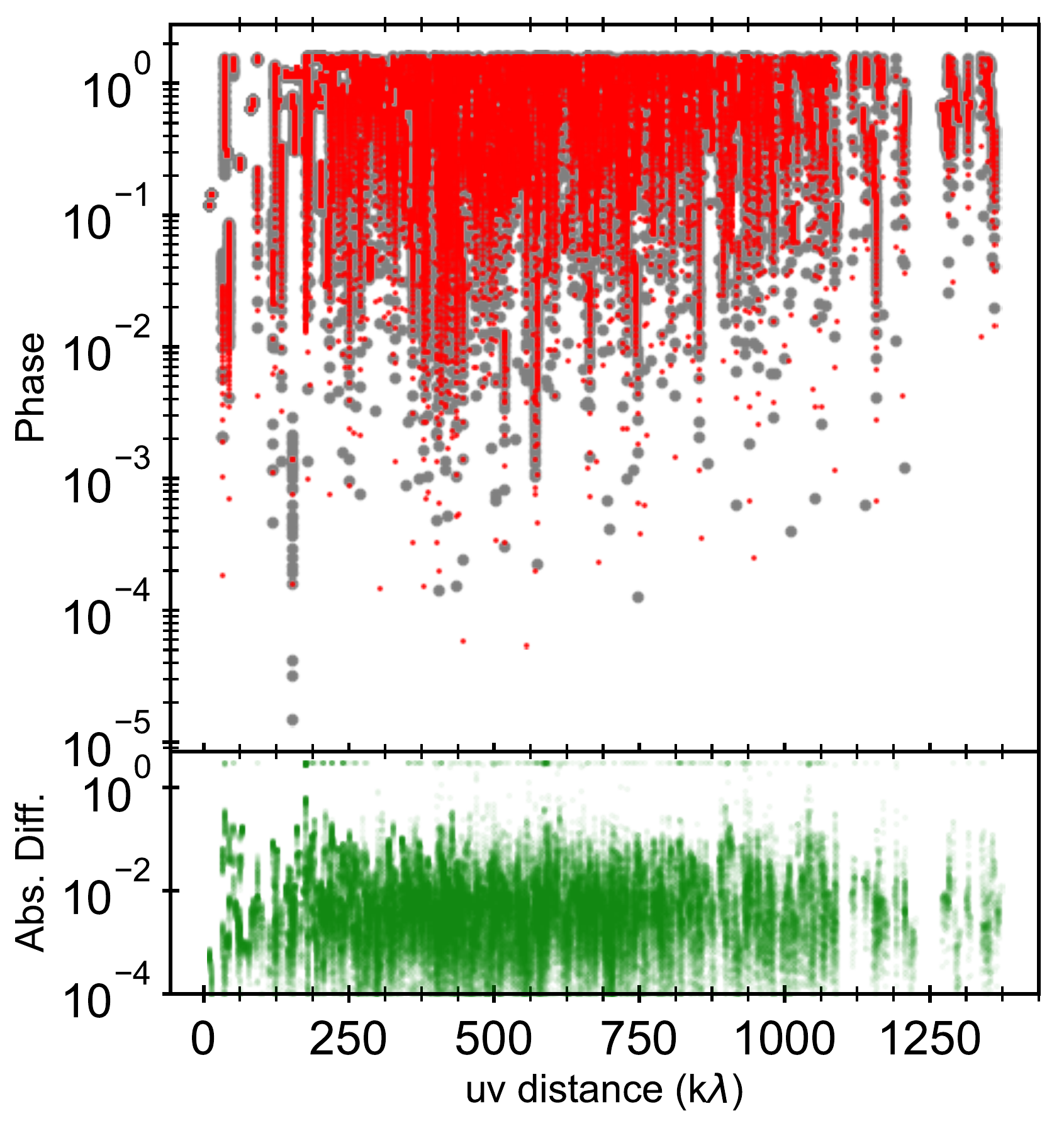}}\\
\caption{ 
Results of the accuracy checks that we carried out against the NRAO CASA package. \textit{Left column:} images of the simulated sources. \textit{Central and right columns:} the comparison of the amplitude and phase of the synthetic visibilities. Each dot represents a \uv-point. The lower panels of the amplitude and phase plots represent, respectively, the relative and the absolute difference between the computed quantities. \textit{First row:} the results from CASA (gray dots) and \galario (red dots) are compared to the analytic solution (yellow dots). \textit{Second and third row:} the results of CASA are compared to those of \galario.}
\label{fig:app.accuracy}
\end{figure*}
}

{
\section{Python reference functions}
\label{app:python.functions}
For completeness, in Figure~\ref{fig:code.pychi2Image} we report the implementation of \func{py\_chi2Image} and \func{py\_chi2Profile}, the Python version of \galario's \func{chi2Image} and \func{chi2Profile} functions. These Python functions are used as the reference for the speed-up factors computed in Section~\ref{sec:performance}. For all the computate-heavy operations they employ only optimized \comm{numpy} and \comm{scipy} functions, as provided in the Anaconda Python distribution. The \func{py\_chi2Image} and \func{py\_chi2Profile} functions are also provided in the unit tests in the online repository.
\begin{figure*}
\begin{minted}[linenos, frame=lines, fontsize=\footnotesize, baselinestretch=0.8]{python}
import numpy as np
from scipy.interpolate import RectBivariateSpline, interp1d

def py_chi2Image(image, dxy, u, v, vis_obs_re, vis_obs_im, weights, dRA=0., dDec=0., PA=0.):
    """ Python implementation of galario `chi2Image` function. """
    nxy = reference_image.shape[0]
    dRA *= 2.*np.pi
    dDec *= 2.*np.pi
    du = 1. / (nxy*dxy)

    # Real to Complex transform
    fft_r2c_shifted = np.fft.fftshift(np.fft.rfft2(np.fft.fftshift(image)), axes=0)

    # rotate (u, v) point coordinates
    cos_PA = np.cos(PA)
    sin_PA = np.sin(PA)
    urot = u * cos_PA - v * sin_PA
    vrot = u * sin_PA + v * cos_PA
    dRArot = dRA * cos_PA - dDec * sin_PA
    dDecrot = dRA * sin_PA + dDec * cos_PA

    # compute interpolation indices
    urot_idx = np.abs(urot)/du
    vrot_idx = nxy/2. + vrot/du
    uneg = urot < 0.
    vroti[uneg] = nxy/2 - vrot[uneg]/du

    # coordinates of FT matrix
    u_axis = np.linspace(0., nxy // 2, nxy // 2 + 1)
    v_axis = np.linspace(0., nxy - 1, nxy)
	
    # sample the Fourier Transform in the (u, v) points
    f_re = RectBivariateSpline(v_axis, u_axis, fft_r2c_shifted.real, kx=1, ky=1, s=0)
    ReInt = f_re.ev(vrot_idx, urot_idx)
    f_im = RectBivariateSpline(v_axis, u_axis, fft_r2c_shifted.imag, kx=1, ky=1, s=0)
    ImInt = f_im.ev(vrot_idx, urot_idx)
    ImInt[uneg] *= -1.  # correct for Real to Complex transform frequency mapping

    # apply the phase change to translate image by (dRA, dDec)
    theta = urot*dRArot + vrot*dDecrot
    vis = (ReInt + 1j*ImInt) * (np.cos(theta) + 1j*np.sin(theta))
    
    chi2 = np.sum(((vis.real - vis_obs_re)**2. + (vis.imag - vis_obs_im)**2.)*weights)

    return chi2



def py_chi2Profile(intensity, Rmin, dR, nxy, dxy, u, v, vis_obs_re, vis_obs_im, weights, dRA=0., dDec=0., PA=0, inc=0.):
    """ Python implementation of galario `chi2Profile` function. """
    inc_cos = np.cos(inc)
    nrad = len(intensity)
    gridrad = np.linspace(Rmin, Rmin + dR * (nrad - 1), nrad)
    ncol, nrow = nxy, nxy
    
    # create the mesh grid
    x = (np.linspace(0.5, -0.5 + 1./float(ncol), ncol)) * dxy * ncol
    y = (np.linspace(0.5, -0.5 + 1./float(nrow), nrow)) * dxy * nrow
    x_axis, y_axis = np.meshgrid(x / inc_cos, y)
    x_meshgrid = np.sqrt(x_axis ** 2. + y_axis ** 2.)

    # bilinear interpolation on the 2d grid to create the image
    intensity *= dxy**2.  # convert to Jansky
    f = interp1d(gridrad, intensity, kind='linear', fill_value=0.,
                 bounds_error=False, assume_sorted=True)
    intensmap = f(x_meshgrid)
    f_center = interp1d(gridrad, intensity, kind='linear', fill_value='extrapolate',
                        bounds_error=False, assume_sorted=True)
    intensmap[int(nrow/2), int(ncol/2)] = f_center(0.)

	# use py_chi2Image to compute the chi square from the image
    chi2 = py_chi2Image(intensmap, dxy, u, v, vis_obs_re, vis_obs_im, weights, dRA=dRA, dDec=dDec, PA=PA)
    
    return chi2
\end{minted}
\label{fig:code.pychi2Image}
\caption{ 
Implementation of \func{py\_chi2Image} and \func{py\_chi2Profile}, the Python version of \galario's \func{chi2Image} and \func{chi2Profile} functions. These Python functions are used as reference for the speed-up factors computed in Section~\ref{sec:performance}.}
\end{figure*}
}

\bsp	
\label{lastpage}
\end{document}